\pdfoutput=1
\documentclass[a4paper,fleqn,usenatbib]{mnras}

\usepackage{newtxtext,newtxmath}
\usepackage{gensymb}
\usepackage{verbatim}

\usepackage[T1]{fontenc}
\usepackage{ae,aecompl}

\usepackage{graphicx}	% Including figure files
\usepackage{amsmath}	% Advanced maths commands

\usepackage{amssymb}	% Extra maths symbols
\usepackage{cancel}
\usepackage{ulem}

\title[Breakup of a synchronous binary system]{Breakup of the Synchronous State of Binary Asteroid Systems}

\author[Hai-Shuo Wang and Xi-Yun Hou]{
Hai-Shuo Wang,$^{1,2,3}$
Xi-Yun Hou,$^{1,2,3}$\thanks{E-mail: houxiyun@nju.edu.cn}
\\
$^{1}$School of Astronomy and Space Science, Nanjing University, Nanjing 210093, China\\
$^{2}$Institute of Space Environment and Astronautics, Nanjing University, Nanjing 210093, China\\
$^{3}$Key Laboratory of Modern Astronomy and Astrophysics, Nanjing University, Ministry of Education, Nanjing 210093, China
}

\date{Accepted XXX. Received YYY; in original form ZZZ}

\pubyear{2020}

\begin{document}
\label{firstpage}
\pagerange{\pageref{firstpage}--\pageref{lastpage}}
\maketitle

\begin{abstract}
	This paper continues the authors' previous work and presents a coplanar averaged ellipsoid-ellipsoid model of synchronous binary asteroid system (BAS) plus thermal and tidal effects. Using this model, we analyze the breakup mechanism of the synchronous BAS. Different from the classical spin-orbit coupling model which neglects the rotational motion's influence on the orbital motion, our model considers simultaneously the orbital motion and the rotational motions. Our findings are following. (1) Stable region of the secondary's synchronous state is mainly up to the secondary's shape. The primary's shape has little influence on it. (2) The stable region shrinks continuously with the increasing value of the secondary's shape parameter $a_B/b_B$. Beyond the value of $a_B/b_B=\sqrt{2}$, the planar stable region for the secondary's synchronous rotation is small but not zero. (3) Considering the BYORP torque, our model shows agreement with the 1-degree of freedom adiabatic invariance theory in the outwards migration process, but an obvious difference in the inwards migration process. In particular, our studies show that the so-called 'long-term' stable equilibrium between the BYORP torque and the tidal torque is never a real equilibrium state, although the binary asteroid system can be captured in this state for quite a long time. (4) In case that the primary's angular velocity gradually reduces due to the YORP effect, the secondary's synchronous state may be broken when the primary's rotational motion crosses some major spin-orbit resonances.
\end{abstract}


%%%%%%%%%%%%%%%%%%%%%%%%%%%%%%%%%%%%%%%%%%%%%%%%%%

%%%%%%%%%%%%%%%%% BODY OF PAPER %%%%%%%%%%%%%%%%%%

\section{Introduction}
\label{sec:Introduction}

	The current long-term evolution picture for the synchronous binary asteroid system (BAS) can be roughly described as follows. In the outwards migration case, it may lead to rapid separation. When the mutual orbit grows to the Hill radius, they may become unbounded due to the third-body perturbation from the Sun and become asteroid pairs \citep{cuk2010orbital,Walsh2015Formation,McMahon2010Secular}. It is also possible that the synchronous state of the secondary is broken before the separation of the BAS reaches the Hill radius, forming the so-called wide asynchronous BAS \citep{Jacobson2014formation}. In the inwards migration case, it may lead to gentle collisions and contact binaries \citep{Scheeres2007Rotational,Jacobson2011dynamics}, or alternatively, the BAS could be trapped in a long-term equilibrium \citep{Jacobson2011Long}.

	Several mechanisms account for the BAS’s evolution. (1) The non-spherical terms of the two asteroids usually cause strong spin-orbit coupling in the BAS due to the close mutual distance and the asteroids’ irregular shapes \citep{Scheeres2002Stability,Walsh2015Formation,Hou2017A}. This mechanism usually cause short-period effects, but may generate long-term effects when influenced by spin-orbit resonances. (2) For the BAS formed by rotational fission, the two asteroids are close to each other. After surviving the violent dynamic stage \citep{Jacobson2011dynamics}, if the secondary is close to the priamry, the tidal effects quickly dissipates energy and usually the BAS quickly goes to the synchronous state. If the secondary has a comparable mass as the primary, the BAS quickly goes to the doubly synchronous state. The strength of the tidal torque quickly decreases with the increase of mutual distance. (3) Once the BAS enters a synchronous or a doubly synchronous state, the BYORP torque begins to work \citep{Cuk2005Effects}. The direction and strength of the torque depends on the asteroids’ shape and orbit. In case it expands the orbit, the BAS may eventually break apart forming asteroid pairs \citep{McMahon2010Secular}, or the synchronous state is broken forming asynchronous BAS \citep{Jacobson2014formation}.In case it shrinks the orbit, the BAS may collapse or enter a long-term equilibrium state with the tidal torque \citep{Jacobson2011Long}. The above three mechanisms will be considered in our work. Except these three mechanisms, the major body flyby and the secular resonances may also play their roles in shaping the BAS’s evoluton route \citep{Fang2012Encounters,Fang2012Kozai}, but, they are not considered in this study.

	About one half of the BASs discovered till now are synchronous BAS \citep{Pravec2016binary}. A natural question is the how large the stable region of the synchrous state is for a synchronous BAS. In a previous paper by the authors \citep{wang2020secondary}, using the sphere-ellipsoid (primary-secondary) model, this problem is studied via the approach of periodic orbits. In the body-fixed frame of the ellipsoid secondary, the synchronous rotation of the secondary can be interpretated as quasi-periodic orbits of the primary in this frame. The quasi-periodic motions are composed of the long-period compoenent and the short-period compoenent, which are closely related with the free librtion amplitude of the synchronous state and the orbit eccentricity respectively. By using this relationship and by describing the stable region in the space spanned by the long-period component and the short-period component, we are able to obtain an anti-correlation between the maximum free libration component and the orbit eccentricity, which is supported by observations. In the first part of this study, we extend the sphere-ellipsoid model to the averaged ellipsoid-ellpsoid model. Our studies show that the primary’s non-spherical terms have neglegible effects on the secondary’s synchronous rotation if the primary rotates fast enough.

	After describing the stable region of the synchronous state, another question is whether the secondary’s rotation can get out of the stable region, i.e., break up during the long-term evolution process. Generally, the synchronous BAS migrates outwards or inwards when influenced by the tidal effects and the BYORP effect. In the case of outwards migration, neglecting the secondary’s rotational motion’s influence on the orbital migration process, and using the adiabatic invariance for the 1-degree of freedom (DOF) model of the 1:1 spin-orbit resonance, \cite{Jacobson2014formation} proposed a mechanism for the formation of some wide asynchronous BASs with small orbit eccentricities. In this study we describe the outwards migration process using the full model, i.e., simultaneously considering the orbital and the secondary’s rotational motions. Our model shows agreement with the adiabatic invariance theory that the free libration amplitude increases and the mutual orbit eccentricity decreases. In case of the inwards migration, a natural result when applying the 1-DOF invariance model is that the free libration amplitude should shrink. However, our full model describes a different evolution picture. The free libration amplitude of the BAS also gradually increases in the inwards migration process and the synchronous state of the BAS breaks up when the free libration amplitude exceeds the stable region. We attribute this phenomenon to the stronger spin-orbit coupling when the two asteroids are closer to each other, which makes the 1-DOF model no longer valid anymore. In particular, we find that the long-term equilbrium state between the tidal torque and the BYORP torque in the inwards migration process is actually a temporary phenomenon. The synchronous state of the BAS eventually breaks up, so does the long-term equilibrium equilibrium state.

	Except the migration process, another mechanism which may cause the breakup of the synchronous state of the BAS is the primary’s spin-orbit resonances. Generally, the primary rotates fast so its rotational motion has no influence on the secondary’s synchronous state. However, when the primary crosses some major spin-orbit resonances, it may cause the break-up of the secondary’s synchronous state due to direct \citep{Batygin2015Spin,Nadoushan2016Geography} and indirect \citep{Hou2017A} spin-spin and spin-orbit-spin coupling mechanism in the BAS. At the end of this study, we present a numerical example showing this phenomenon.

	In total, two mechanisms that can cause the breakup of the synchronous state of the BAS are studied, the migration process and the primary’s spin-orbit resonances. The findings of this study are fourfolds:

	(1) The stable region of the secondary's synchronous state is mainly determined by the secondary's shape. On the other hand, the primary's shape has little influence on it, as long as the primary rotates fast enough. 
	
	(2) With the increase of $a_B/b_B$, stable region shrinks continuously. When $a_B/b_B$ is larger than $\sqrt{2}$, the planar stable region for the secondary's synchronous state still exists, with an upper limit for the orbit eccentricity no larger than 0.05.

	(3) The classical model which neglects rotational motions' influence on the orbital motion is a good approximation in the outwards migration but not in the inwards migration. Also, we find that the long-term equilibrium state between the BYORP torque and the tidal torque will eventually break apart in our full model.

	(4) Although the primary's shape has little influence on the secondary's synchronous state, due to the strong spin-orbit coupling \citep{Hou2017A,Hou2018note}, the situation may change when the primary's rotational speed gradually reduces and crosses some major spin-orbit resonances. The secondary's synchronous state may be broken during the primary's spin-orbit resonance crossing. 

	This paper is organized as follows. Sec.\ref{sec:ModelDescription} introduces the dynamical model used in this study. Sec.\ref{sec:DynamicsOfTheAveragedSystem} describes the dynamics of an averaged ellipsoid-ellipsoid model. A method of describing the stable region of the secondary's synchronous state is introduced in this section. Sec.\ref{sec:BreakupMechanismI} studies the break-up mechanism of the synchronous state due to the outwards/inwards migration process. Sec.\ref{sec:BreakupMechanismII} studies the break-up mechanism of the synchronous state due to the primary's spin-orbit resonances. Section 6 concludes the study.
 
\section{Model Description}
\label{sec:ModelDescription}

	The dynamical model used in this article is described in this section. The current work is a continuation of the authors' previous work \citep{wang2020secondary} which studied the correlation between the mutual orbit eccentricity and secondary's synchronous rotation state. Compared with the previous work, there are two differences in the force model used in the current work: 
	(1) Tidal torque and BYORP torque are considered, which cause the migration of the binary asteroid system (BAS); 
	(2) Non-spherical terms of the primary are also considered. 
	In the following, some details on the forced model used in the current work will be given. 

	\subsection{Planar Two-ellipsoid Model}
	\label{sec:Planar2ElliposoidModel}
\begin{figure}
			\includegraphics[width=\columnwidth]{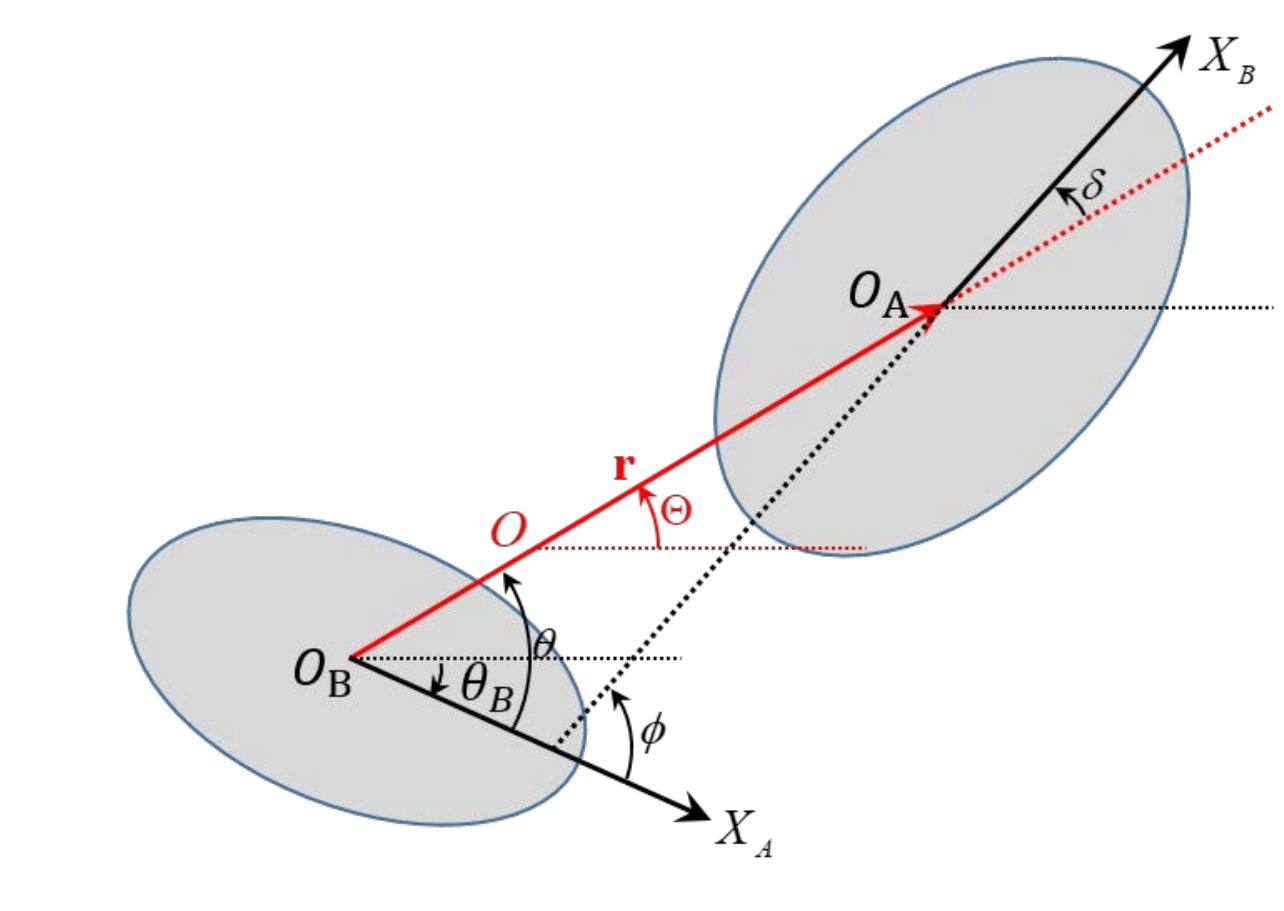}
			\caption{Relative geometry of the planar two-ellipsoid model for the binary asteroid system}
			\label{fig:ModelAB}
		\end{figure}

	Two asteroids in the BAS are approximated as two tri-axial ellipsoids. Moreover, we assume that the two asteroids are rotating along their shortest axis and their mutual orbital planes coincides with their rotation plane. This is the well-known planar two-ellipsoid model used in many previous studies \citep{Ureche1974Ellipsoid,Ureche1978Elliposid,Bellerose2008Energy,Hou2017A,Hou2020Forced,Wang2018Stabilization}. Fig.\ref{fig:ModelAB} shows the relative geometry of BAS and $A$ is the primary body, which is larger than secondary body $B$. The coordinate system $O-xy$ is an inertial frame and the coordinate system $O_*-X_*Y_*$ (the subscripts $*$ means $A$ or $B$) is the body-fixed frame of $A$ or $B$ in which we define the $X_*$ axis along $A$'s or $B$'s longest axis. From the above definition, the angles showed in Fig.\ref{fig:ModelAB} satisfy
		\begin{equation}
			\Theta = \theta + \theta_B, \qquad \phi = \theta_A-\theta_B, \qquad \delta = \theta - \phi.
		\label{eq:AnglesEquation}
		\end{equation}
		The semi-axes and masses of $A$ and $B$ are denoted as
		\begin{equation}
			a_{A}, b_{A}, c_{A}, m_{A} ; \quad a_{B}, b_{B}, c_{B}, m_{B}.
		\label{eq:EllipsoidParameters}
		\end{equation}
		The non-spherical gravity of $A$ or $B$ is truncated at the second order and expressed in the form of spherical harmonics. The coefficients $J_2$ and $J_{22}$ of $A$ and $B$ are given by \citep{Balmino1994gravitational}.
		\begin{equation}
			J_{2}^{*}=\frac{a_{*}^{2}+b_{*}^{2}-2 c_{*}^{2}}{10 \bar{a}_{*}^{2}}, \quad J_{22}^{*}=\frac{a_{*}^{2}-b_{*}^{2}}{20 \bar{a}_{*}^{2}}.
		\label{eq:J2}
		\end{equation}
		The reference radius $\bar{a}_{*}$ is defined as $\bar{a}_{*}=\left(a_{*} b_{*} c_{*}\right)^{1 / 3}$ in this study. Following normalized units are used in this work
		\begin{equation}
			\left\{\begin{array}{l}
			{[M]=m_{A}+m_{B}} \\
			{[L]=\bar{a}_{A} + \bar{a}_{B}} \\
			{[T]=\left([L]^{3} / G[M]\right)^{1 / 2}}
			\end{array}\right.
		\label{eq:NomalizedUnit}
		\end{equation}
		According to \citep{Hou2017A,Hou2018note}, the mutual potential of $A$ and $B$ truncated at second order is
		\begin{equation}
			U=-m\left[\frac{1}{r}+\frac{1}{r^{3}}\left(A_{1}+A_{2} \cos 2 \theta+A_{3} \cos 2 \delta\right)\right],
		\label{eq:Potential}
		\end{equation}
		in which, we define $m_A$ and $m_B$ as the masses of the two ellipsoids, respectively. The parameter $\mu$ is defined as $\mu=\frac{m_{B}}{m_{A}+m_{B}}$, and $m$ is defined as $m = \mu(1-\mu)$. The coefficients $A_i (i=1,2,3)$ are defined as
		\begin{equation}
			\left\{\begin{array}{l}
			{A_{1}=\frac{1}{2}\left(J_{2}^{A} \alpha_{A}^{2}+J_{2}^{B} \alpha_{B}^{2}\right)} \\
			{A_{2}=3 J_{22}^{B} \alpha_{B}^{2}} \\
			{A_{3}=3 J_{22}^{A} \alpha_{A}^{2}} \\
			{\alpha_{*}=\bar{a}_{*} / a}.
			\end{array}\right.
		\label{eq:Letters}
		\end{equation}
		Equation of motion (EOM) for the BAS truncated at the second order of the mutual potential is given as \citep{Hou2017A}:
		\begin{equation}
			\left\{
				\begin{array}{lr}
					{\ddot{r} = r\dot{\Theta}^2-\frac{1}{r^2}-\frac{3}{r^4}(A_1+A_2\cos(2\theta)+A_3\cos(2\delta))}\\
					{\ddot{\Theta} = -2\frac{\dot{r}}{r}\dot{\Theta}-\frac{2}{r^5}[A_2\sin(2\Theta-2\theta_B)+A_3\sin(2\Theta-2\theta_A)]}\\
					{\ddot{\theta}_A = \frac{2mA_3}{I_z^A}\frac{\sin(2\Theta-2\theta_A)}{r^3}}\\
					{\ddot{\theta}_B = \frac{2mA_2}{I_z^B}\frac{\sin(2\Theta-2\theta_B)}{r^3}}
				\end{array}
			\right.
		\label{eq:YHC4}
		\end{equation}

	\subsection{Tidal Torque}
	\label{sec:TidalTorque}

		The evolutionary consequences of mutual body tides have been considered for the BAS since the discovery of the first asteroid satellite, Dactyl, about (243) Ida \citep{Petit1997long,Hurford1999tidal}. Due to the difference between roational rate and orbital rate, tidal torque is produced in evolution. This torque dissipates system energy in the form of heat and transfers momentum between asteroids and mutual orbit, while the total angular momentum is conserved within the BAS. Different from the Earth-Moon system, where the most of energy is dissipated at the ocean-seabed interface and in deep ocean  itself \citep{Egbert2000significant}, the tidal torque in the BAS dissipates energy throughout the solid body tide. Some new results indicate that tidal dissipation in rubble pile asteroids may be much stronger \citep{Scheirich2015binary} than previously expected \citep{Goldreich2009tidal}, so a remarkable influence on the evolution of the BAS by tidal torque is expected.	

		Binary near-Earth asteroids are good test candidates for the tidal effect. In the leading hypothesized formation mechanism, BASs have a rapidly rotating primary and a secondary that is quickly tidally locked after rotational disruption. \citep{Jacobson2011dynamics,Walsh2008steady,Scheeres2007Rotational}. The overall effect of all tides on mutual orbit is that the semimajor axis expands and the eccentricity damps. According to the classical tidal theory \citep{Murray1999solar}, the tidal torque of one body $j$ acting on the other body $i$ is:
		$$\Gamma_{i}=\operatorname{sign}\left(\dot{\phi}_{i}\right) \frac{3}{2} k_2^i\left(\frac{3}{4 \pi \rho_{i}}\right)^{2} \frac{G M_{i}^{2} M_{j}^{2}}{r_{i j}^{6} R_{i}} \sin \left(2 \epsilon_{i}\right)$$
		where $k_2^i$ is the tidal Love number of the body $i$ and $\epsilon_i$ is the tidal lag angle. $R_{i}$ is the radius of body $i$. The sign of $\dot{\phi}_{i}$ determines whether the tidal bulge is leading or trailing the tide-raising satellite, which determines the direction of angular momentum transfer between the orbit and the spin state. The tidal lag angle can be related to the specific tidal dissipation function $Q$ which describes how effective the body is at tidally dissipation of energy \citep{Murray1999solar}
		$$Q=\frac{1}{\tan 2 \epsilon} \approx \frac{1}{2 \epsilon}$$

		However, this classical torque presents a problem. When the angular velocity $\dot{\phi}_{i}$ crosses zero, torque direction changes instantaneously. It is unphysical and also leads to a difficulty in numercial propagation. Therefore, a smooth transition should be considered near to zero. To overcome this problem, following \citep{Jacobson2011dynamics}, we modify the tidal torque as,
		\begin{equation}
			\ddot{\theta}_*^{tid} = 
			\left\{
				\begin{array}{lr}
					sgn(\dot{\Theta}-\dot{\theta}_*)\frac{3k_2^*\mu^2}{2Q_*r^6}\frac{\alpha_*^5}{I_z^*}  \quad |\dot{\Theta}-\dot{\theta}_*|>\delta_*\\
					(\dot{\Theta}-\dot{\theta}_*)\sqrt{\frac{\pi \rho_*}{6}} \qquad |\dot{\Theta}-\dot{\theta}_*|\le\delta_* \\
				\end{array}
			\right.
		\label{eq:Tides}
		\end{equation}
		where, $$\delta_* = \ddot{\theta}_*^{tid}\sqrt{\frac{6}{\rho_*\pi}}.$$
		The notation $\rho_*$ is normalized density of the body $A$ or $B$. Following \citep{Jacobson2011Long} which estimates the magnitude of $Q/k_2$ according to the long-term balance between the tidal torque and the BYORP torque, we take following values throughout computing the tidal torque.
		\begin{equation}
			Q_*/k^*_2 = 6\cdot10^5\frac{r_*}{1km}\\
		\label{con:ConstTidal}
		\end{equation}
		in which $r_*$ is the radius of the asteroid which takes the value of the reference radius $\bar{a}_{*}$ in this work. 

		According to the law of conservation of the total angular momentum, the torque on the mutual orbit caused by the tidal effects is ,
		\begin{equation}
			F t=\frac{1}{m r}\left[I_{z}^{A} \ddot{\theta}_{A}^{t i d}+I_{z}^{B} \ddot{\theta}_{B}^{t i d}\right].
		\label{eq:TraverseTidalForce}
		\end{equation} 
		In this torque, notice that,
		\begin{equation}
			\left|\frac{I_{z}^{A} \ddot{\theta}_{A}^{t i d}}{I_{Z}^{B} \ddot{\theta}_{B}^{t i d}}\right|=\frac{r_{B}}{r_{A}}\left(\frac{\mu_{B}}{\mu_{A}}\right)^{2}\left(\frac{\alpha_{A}}{\alpha_{B}}\right)^{5}=\left(\frac{r_{B}}{r_{A}}\right)^{2}<1.
		\label{eq:TidalTorqeInequation}
		\end{equation}
		This means that the tidal torque on $B$ generated by $A$ is always larger than the tidal torque on $A$ generated by $B$. Also notice that $\ddot{\theta}_B^{tid}$ equals zero if $B$ is tidally locked (i.e. $\dot{\Theta}=\dot{\theta}_B$). $\ddot{\theta}_B^{tid}$ and $\ddot{\theta}_A^{tid}$ are both equal to zero when the BAS enters a doubly synchronous state (i.e. $\dot{\Theta}=\dot{\theta}_B$ and $\dot{\Theta}=\dot{\theta}_A$). 

		Changes in the semimajor axis and eccentricity can easily be deduced as \citep{Cuk2005Effects}
		\begin{equation}
			\left\{\begin{array}{l}
			<\dot{a}>\simeq \frac{2 \sqrt{1-e^{2}}}{3^{n}} F_t \\
			<\dot{e}>\simeq-\frac{3 e}{2 n a} F_t
			\end{array}\right..
		\label{eq:TideElementsEvolution}
		\end{equation}

	\subsection{BYORP Torque}
	\label{sec:BYORPTorque}
		Apart from the tidal torque, for a synchronous or doubly synchronous BAS, the thermal BYORP effect plays its role. Instead of modifying the spin state of an asteroid, the BYORP effect modifies the mutual orbit. Readers can refer to \citep{Cuk2005Effects,Mcmahon2010detailed} for physical interpretation and calculation of the BYORP torque. This effect, first suggested by \citep{Cuk2005Effects}, evolves the orbit of the binary on fairly fast timescales (in less than $10 ^ 5$ years for a primary of $1 km$ and secondary of $0.3 km$ at $1 AU$). According to \citep{Jacobson2014formation}, the synchronous state of the secondary can be broken by the outwards migration process, forming the so-called wide asynchronous BAS. In this work, we will describe how the BAS migrates outwards in the phase space. Different from \citep{Jacobson2014formation}, the migration process is displayed in a 2-DOF model. 

		By averaging over the mutual orbit of the binary system and its orbit around Sun, \citep{Mcmahon2010detailed} gave detailed formulae for the long-term effects on the orbit elements. Instead of using these detailed formulae, we use the simple formula given by \citep{Steinberg2011binary} to describe the torque on the mutual orbit generated by the BYORP effect
		\begin{equation}
			\ddot{\Theta}^{B Y}=\frac{B_{s} \pi \alpha_{B}^{2}}{\left(a_{s} /[L]\right)^{2} \sqrt{1-e_{s}^{2}}} f_{B Y} \frac{\overline{F_{s}}}{m r}
		\label{eq:BYORP}
		\end{equation}
		where, $B_s$ is the Lambertian scattering coefficient which takes the value of $2/3$ in this work. $F_s$ is the solar radiation constant and takes the value of $10^{17} kgm/s^2$ and $\overline{F_{s}}$ is the normalized $F_s$ formulated as $\overline{F_{s}} = F_s [T]^2/([M][L])$. $a_s(e_s)$ is the semi-major axis (eccentricity) of the asteroid's orbit with respect to the Sun and takes the value of 1AU(zero) in this work. $|f_{BY}|$ is the BYORP coefficient, which is usually a small quantity, and takes the average value of 0.01 given by \citep{Steinberg2011binary}, either positive for expanding or negative for shrinking the mutual orbit. $\alpha_{B}$ was defined in Eq.\ref{eq:Letters}. In our model, we assume that the rotational states of $A$ or $B$ are not directly influenced caused by the BYORP effect.

		For a circular mutual orbit, Eq.\ref{eq:BYORP} is identical to the result by applying a constant tangential torque to the orbit. This vivid physical explanation is actually inappropriate because actually there is no such a force.The BYORP effect is the averaged thermal effect on the orbital elements \citep{Mcmahon2010detailed}. Nevertheless, to simplify our model, we directly add the BYORP torque given by Eq.\ref{eq:BYORP} to the second equation of Eq.\ref{eq:YHC4}.
		It can be easily shown that the secular evolution of the mutual orbit’s semi-major axis $a$ and eccentricity $e$ is,
		\begin{equation}
			\left\{\begin{array}{l}
			<\dot{a}>\simeq \frac{2 \sqrt{1-e^{2}}}{n} F_B \\
			<\dot{e}>\simeq-\frac{3 e}{2 n a} F_B
			\end{array}\right.,
		\label{eq:BYORPElementsEvolution}
		\end{equation}
		from which we know
		\begin{equation}
			\frac{<\dot{a}>}{<\dot{e}>}=-\frac{4 a}{e} \frac{1}{1-e^{2}} \sim-\frac{4 a}{e}.
		\label{eq:OppositeEvolution}
		\end{equation}
		This means the orbit semi-major axis evolves in the opposite direction to the orbit eccentricity, i.e., the averaged orbit eccentricity decreases if the orbit migrates outwards and increases if the orbit migrates inwards. We will verify this phenomenon in following numerical simulations.

	\subsection{Averaged EOM}
	\label{sec:AveragedEOM}
	In the synchronous BAS, the primary usually rotates much faster than the secondary. Since the primary is usually larger than the secondary, its rotational state is little influenced by the mutual orbit and the secondary's rotation. As a result, unless we focus on the rotational effect of the primary, we can actually average the equations of motion over the primary's rotation. Eq.\ref{eq:YHC4} after averaging becomes 
	\begin{equation}
		\left\{\begin{array}{l}
		\ddot{r}=r \dot{\theta+\theta_B}^{2}-\frac{1}{r^{2}}-\frac{3}{r^{4}}[A_{1}+A_{2} \cos (2 \theta)] \\
		\ddot{\theta}=-2 \frac{\dot{r}}{r} (\theta+\theta_B)-\frac{2}{r^{5}}A_{2} \sin \left(2 \theta\right)\\
		\ddot{\theta}_{B}=\frac{2m A_{2}}{I_{z}^{B}} \frac{\sin \left(2 \theta\right)}{r^{3}}
		\end{array}.\right.
	\label{eq:AveragedEOM_bodyfixed}
	\end{equation}
	This equation has the same form as the sphere-ellipsoid model used in our previous work \citep{wang2020secondary}, but the coefficients contain the averaged terms from the primary's gravity.
	$$
	A_{1}=\frac{1}{2}\left(J_{2}^{A} \alpha_{A}^{2}+J_{2}^{B} \alpha_{B}^{2}\right) 
	\qquad
	A_{2}=3 J_{22}^{B} \alpha_{B}^{2}
	\qquad
	A_3 = 0.
	$$
	To distinguish from the sphere-ellipsoid model, hereafter in this paper, we call the averaged Eq.\ref{eq:AveragedEOM_bodyfixed} as the averaged ellipsoid-ellipsoid model. If the primary $A$ is a sphere, it means $J_{22}^A = 0,J_{2}^{A} = 0$. The averaged ellipsoid-ellipsoid model degenerates to a sphere-ellipsoid model with $A_{1}=\frac{1}{2}J_{2}^{B} \alpha_{B}^{2}$ and $A_3 = 0$.By adding the long-term tidal effect and the BYORP effect to Eq.\ref{eq:AveragedEOM_bodyfixed}, we have

	\begin{equation}
		\left\{\begin{array}{l}
		\ddot{r}=r \dot{\Theta}^{2}-\frac{1}{r^{2}}-\frac{3}{r^{4}}[A_{1}+A_{2} \cos (2 \theta)] \\
		\ddot{\Theta}=-2 \frac{\dot{r}}{r} \Theta-\frac{2}{r^{5}}A_{2} \sin \left(2 \Theta-2 \theta_{B}\right)\\
		\qquad -\frac{1}{m r^{2}}\left[I_{z}^{A} \ddot{\theta}_{A}^{t i d}+I_{z}^{B} \ddot{\theta}_{B}^{t i d}\right] \pm \ddot{\Theta}^{B Y} \\
		\ddot{\theta}_{B}=\frac{2m A_{2}}{I_{z}^{B}} \frac{\sin \left(2 \Theta-2 \theta_{B}\right)}{r^{3}}
		\end{array},\right.
	\label{eq:AveragedEOM}
	\end{equation}
	which can be used to study the long-term effect of the tidal effect and the BOYRP effect on the secondary's rotation.

	In total, three models will be used in this work. Eq.\ref{eq:AveragedEOM_bodyfixed} will be used as a continuation of the previous work, studying the averaged effect on the secondary's rotation by the primary's non-spherical gravity. Eq.\ref{eq:AveragedEOM} will be used to study the long-term effect on the secondary's rotation by the tidal effect and the BYORP effect. Eq.\ref{eq:YHC4} will be used to study the relatively short-term resonant effect on the secondary's rotation by the primary's rotational motion.

\section{Dynamics of the Averaged System}
\label{sec:DynamicsOfTheAveragedSystem}

	As a continuation of the previous work, we study the effect of the primary's gravity on the secondary's synchronous rotational state. Same as the previous work, using the conservation of the total angular momentum
	\begin{equation}
		K = mr^2(\dot{\theta}_B+\dot{\theta}) + I_z^B\dot{\theta}_B,
	\label{eq:TotalMomentum}
	\end{equation}
	we express $\dot{\theta}_B$ as a function of $K$, $\dot{\theta}$, $\dot{\theta}_A$ and $r$. Substituting this function into Eq.\ref{eq:AveragedEOM_bodyfixed}, we can reduce one degree of freedom of the system, in the form of
	\begin{equation}
		\left\{\begin{array}{l}
		\ddot{r}=r\frac{(I_z^A\dot{\theta} - I_z^B\dot{\theta}-K)^2}{(mr^2 + I_z^B)^2}-\frac{1}{r^{2}}-\frac{3}{r^{4}}\left[A_1 + A_{2} \cos (2 \theta)\right] \\
		\ddot{\theta}=\frac{2\dot{r}(I_z^A \dot{\theta}_A - I_z^B \dot{\theta} - K)}{r(mr^2 + I_z^B)}-\frac{2A_{2} \sin (2 \theta) (I_z^B + mr^2)}{I_{z}^{B} r^{5}}
		\end{array}\right.
	\label{eq:AveragedEOM_K}
	\end{equation}
	As mentioned previously, this equation of motion has the same form as that of the sphere-ellipsoid model in our previous work. The only difference is that the value of the coefficient $A_1$ in this case incorporates the averaged effect of the primary's non-spherical gravity. In the following, we will show the difference between the averaged ellipsoid-ellipsoid model and the sphere-ellipsoid model, in terms of the equilibrium points and stability, periodic orbits, and stable region. 

	\begin{table}
		\centering
		\caption{System parameters adopted for this work. Note that the primary's rotation velocity used in averaged model is a fixed value.}
		\label{tab:SystemParameters}
		\begin{tabular}{lccr} % four columns, alignment for each
			\hline
			Asteroid parameter & symbol & value\\
			\hline
			Mean radii of the primary & $r_A$ & 800$m$ \\
			Mean radii of the secondary & $r_B$ & 450$m$\\
			Primary(shape) elongation  & $a_{A}/b_{A}$ & 1.2\\
			Secondary(shape) elongation	&	$a_{B}/b_{B}$ & 1.2\\
			Bulk density of the asteroids  & $\rho$ & 2100$kg m^{-3}$\\
			Mean mutual orbit distance  & $R$ & 5000$m$\\
			Mass fraction  & $\mu$ & 0.151\\
			Primary's rotation velocity  & $\dot{\theta}_A$ & 1.5\\
			\hline
			BYORP parameter & symbol & value\\
			\hline
			Lambertian scattering coefficient  & $B_s$ & $2/3$\\
			solar radiation constant  & $F_s$ & $10^{17}kgm/s^2$\\
			BYORP coefficient  & $f_{BY}$ & $0.01$\\
			\hline
			Tidal parameter & symbol & value\\
			\hline
			Love number  & $Q_*/k_2^*$  & $6 \cdot 10^5 r_*/1km$\\
			\hline
		\end{tabular}
	\end{table}

	\subsection{Equilibrium Points and Stability}
	\label{sec:EquilibriumPointsAndStability}

\begin{figure}
			\includegraphics[width=\columnwidth]{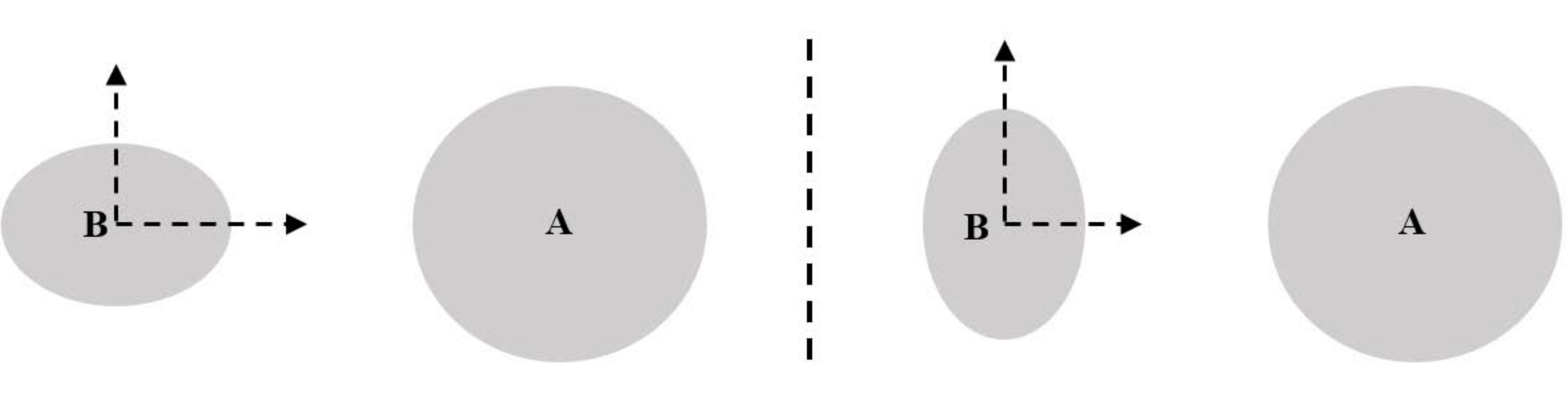}
			\caption{Two configurations of the averaged ellipsoid-ellipsoid system.}
			\label{fig:EquilibriumConfugurtion2}
		\end{figure}

		Same as the sphere-ellipsoid model in our previous work, there are two types of equilibrium points for the averaged system by Eq.\ref{eq:AveragedEOM_K}. Since Eq.\ref{eq:AveragedEOM_K} describes the orbital motion of the primary in the secondary's body-fixed frame, the two types of equilibrium points actually correspond to two configurations of the exact synchronous state of the secondary---the long-axis mode ($\theta=0$, the secondary's long axis always pointing at the primary) and the short-axis mode ($\theta=\pi/2$, the secondary's short axis always pointing at the primary), as illustrated in Fig.\ref{fig:EquilibriumConfugurtion2}.

		The equilibrium points can be obtained by setting all velocities ($\dot{r},\dot{\theta}$) and accelerations ($\ddot{r},\ddot{\theta}$) to zero in Eq.\ref{eq:AveragedEOM_K}. As a result, we have
		\begin{equation}
			\left\{\begin{array}{l}
			\theta = 0,\pi/2,\pi,3\pi/2 \\
			0 = (\frac{I_z^A \dot{\theta}_A - K}{mr^2+I_z^B})^2 - \frac{1}{r^3} - 3\frac{A_1+A_2\cos(2\theta)}{r^5}
			\end{array}\right..
		\label{eq:EPsEquation}
		\end{equation}
		For a fixed value of the mean mutual orbit distance $r$, the total angular momentum $K$ can be solved from the second equation of Eq.\ref{eq:EPsEquation}, or  vice versa, the mean mutual distance $r$ can be solved if the value of $K$ is fixed. We denote the solution as $r_0$ and $\theta_0$. In our work, parameters of the example BAS used are shown in Table \ref{tab:SystemParameters}. According to Table \ref{tab:SystemParameters}, distance between two bodies is $R = 5000m$. If the secondary is trapped in the exact synchronous state at this distance, the dimensionless mutual orbit distance is $r_0 = 3.3333$. The system momentum is $K = 0.4128$ for the long-axis mode ($\theta=0$), or $K = 0.4125$ for the short-axis mode ($\theta=\pi/2$). 

\begin{figure*}
			\begin{minipage}[t]{0.5\linewidth}
				\centering
				\includegraphics[width=\columnwidth]{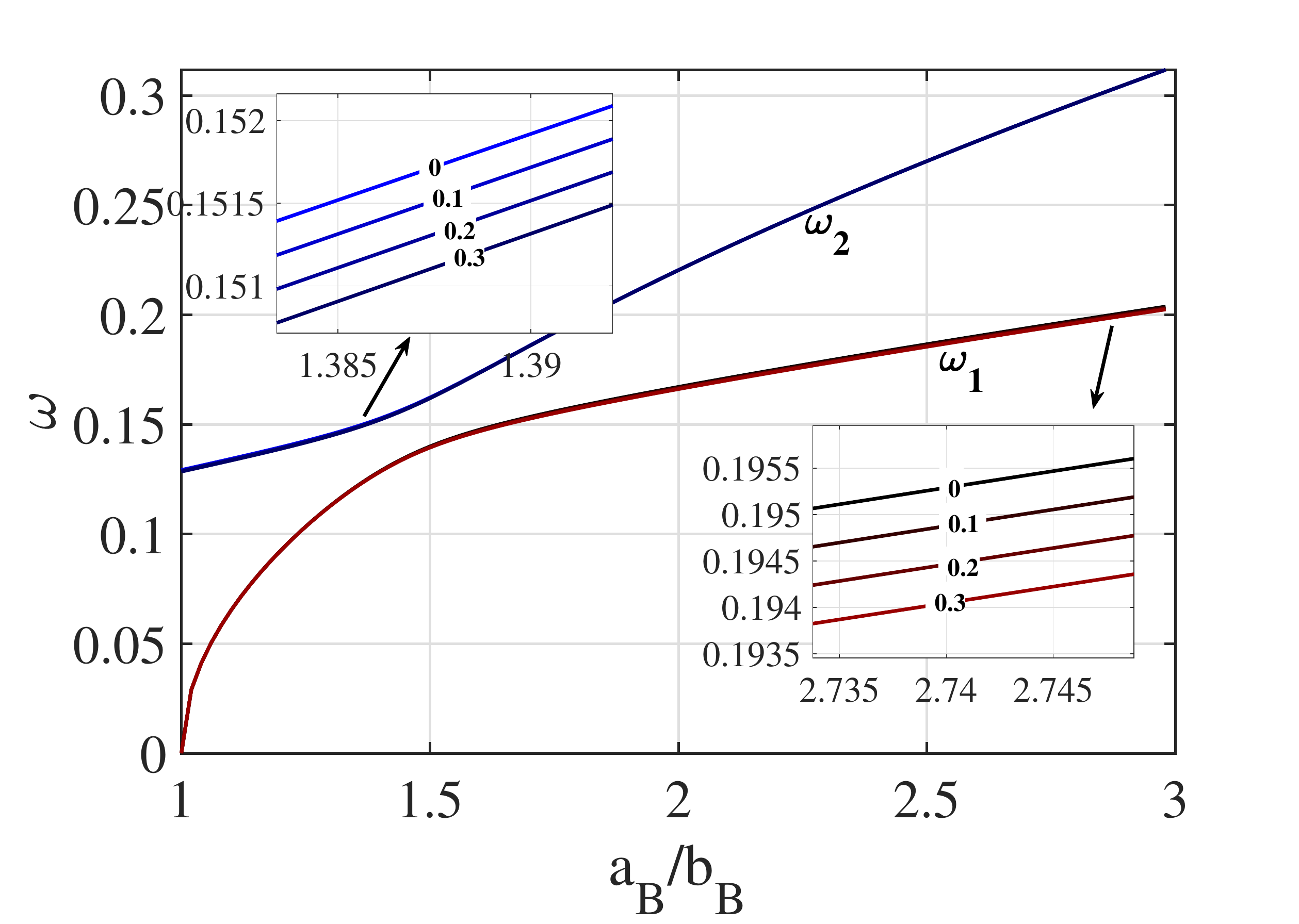}
			\end{minipage}%
			\begin{minipage}[t]{0.5\linewidth}
				\centering
				\includegraphics[width=\columnwidth]{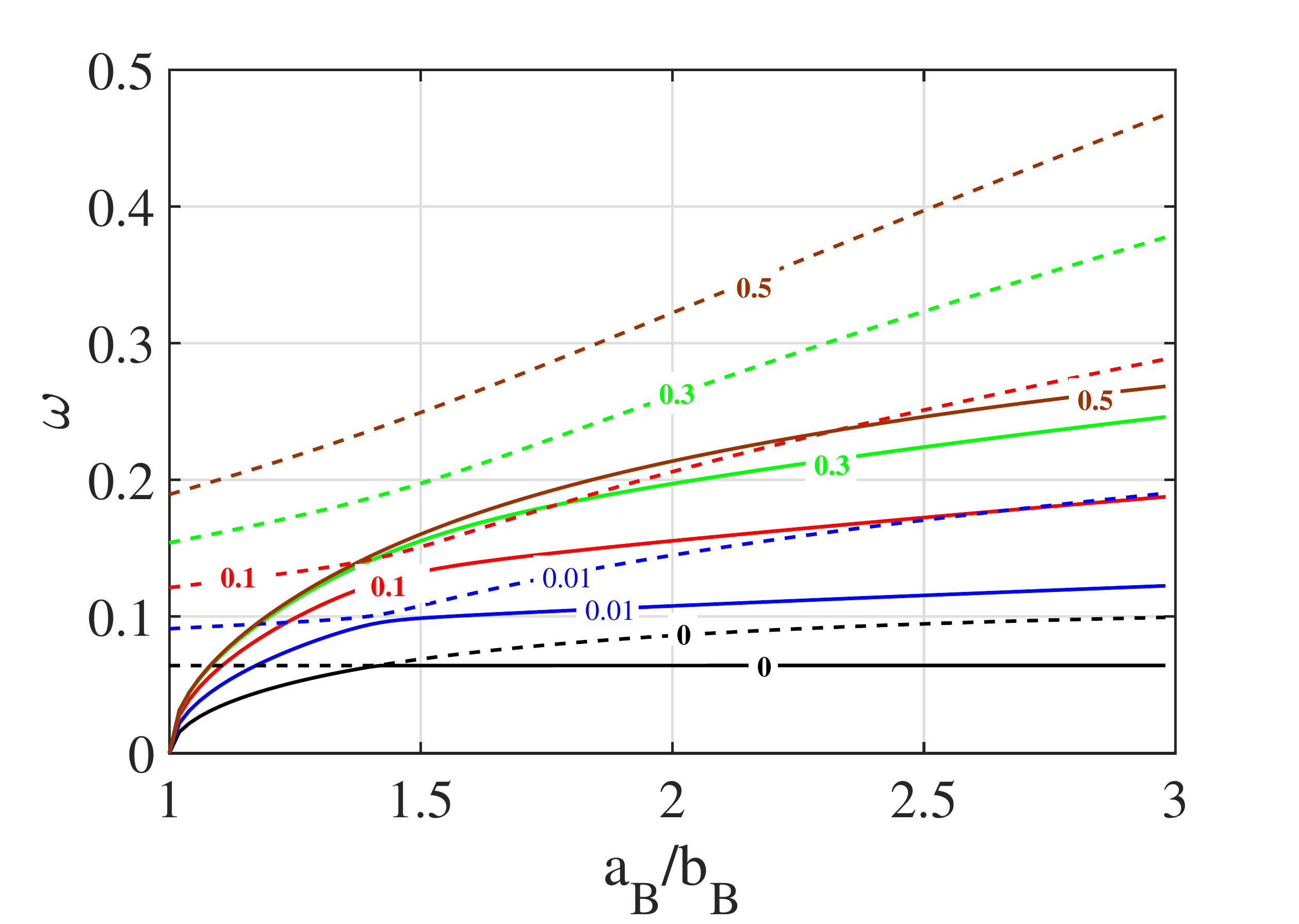}
			\end{minipage}%
			\caption{Curves of two frequencies with respect to the secondary elongation parameter $a_B/b_B$. $\omega_i$ is the frequency of periodic orbits introduced in Sec.\ref{sec:DynamicsOfTheAveragedSystem}. $i = 1,2$ means the frequency of the long-/short- period orbit. The abscissa is secondary's shape parameter $a_B/b_B$ which is less than 1.5 for currently observed synchronous BAS. Different curves in the left frame correspond to different values of the primary's $J_2^A$ term. Boxes are local magnification of the curves, and the values of $J_2^A $ are tagged on the curves. Different curves in the right frame correspond to different values of the mass parameter $\mu$ whose values are tagged on the curves.}
		\label{fig:W0vsPar1}
		\end{figure*}

		Stability of the equilibrium points can be determined by eigenvalues of the variation matrix associated with the equilibrium points. The variational matrix takes the form of
		\begin{equation}
			\mathbf{A_{4 \times 4}}=\left(\begin{array}{cccc}
			0 & 0 & 1 & 0 \\
			0 & 0 & 0 & 1 \\
			a_{31} & a_{32} & 0 & a_{34} \\
			a_{41} & a_{42} & a_{43} & a_{44}
			\end{array}\right)
		\label{eq:A44Matrix}
		\end{equation}
		Detailed expressions for $a_{i j}$ can be found in Appendix \ref{sec:AppendA}. The characteristic equation for this matrix is simplified
		\begin{equation}
			\lambda^{4}-\left(a_{31}+a_{42}+a_{34} a_{43}\right) \lambda^{2}+a_{31} a_{42}=0.
		\label{eq:CharacteristicEquation}
		\end{equation}
		Setting $s_{i}=\lambda^{2},$ roots of Eq.\ref{eq:CharacteristicEquation} are
		\begin{equation}
			\begin{aligned}
			s_{1} &=\frac{\left(a_{31}+a_{42}+a_{34} a_{43}\right)}{2}+\frac{\sqrt{\left(a_{31}+a_{42}+a_{34} a_{43}\right)^{2}-4 a_{31} a_{42}}}{2} \\
			s_{2} &=\frac{\left(a_{31}+a_{42}+a_{34} a_{43}\right)}{2}-\frac{\sqrt{\left(a_{31}+a_{42}+a_{34} a_{43}\right)^{2}-4 a_{31} a_{42}}}{2}
			\end{aligned}
		\label{eq:RootOFEquation}
		\end{equation}
		Stability of the equilibrium points can be different depending on the sign of $s_1$ and $s_2$ (see \citep{wang2020secondary} for more details). For the synchronous BAS in which the primary is usually much larger than the secondary, the long-axis mode is usually stable while the short-axis mode is usually unstable, i.e., the secondary in the synchronous BAS usually has its long axis pointing at the primary. For the stable long-axis mode, we introduce 
		$$\omega_1=\sqrt{-s_1},\quad \omega_2=\sqrt{-s_2}.$$ 
		We assume $\omega_1<\omega_2$, which means $\omega_1$ is the frequency of long-period orbit and $\omega_2$ is the frequency of the short-period orbit. These two frequencies rely on system parameters, for example, the shape and the distance of two bodies. Here, the primary's shape is our focus. In Eq.\ref{eq:J2}, we express this term $J_2^A$ as a function of primary's shape parameters ($a_A,b_A,c_A$). By changing the shape parameters $a_A/b_A$ and $b_A/c_A$, we can get different values of $J_2^A$. By fixing the mass parameter $\mu$, the left frame of Fig.\ref{fig:W0vsPar1} shows how the two basic frequencies are influenced by the primary's $J_2^A$ term. The abscissa is the secondary's shape parameter $a_B/b_B$. Different curves correspond to different values of the primary's $J_2^A$ term. An obvious fact is that the two frequencies of the secondary's synchronous state are little influenced by the primary's shape parameter.

		We take a step further. By fixing $J_2^A=0$, the averaged ellipsoid-ellipsoid model degenerates to the sphere-ellipsoid model. The right frame of Fig.\ref{fig:W0vsPar1} shows how the two basic frequencies change with the secondary's shape parameter and the mass parameter. For the special case of $\mu=0$, the two basic frequencies coincide with each other at the specific value of $a_B/b_B=\sqrt{2}$. As we have pointed out in our previous paper, the $\mu=0$ case of our model is the conventional model for the spin-orbit coupling \citep{Murray1999solar}. In the conventional model, this specific shape parameter is also obtained when the free libration frequency of the synchronous state $\omega_{lib}$ equals the orbital frequency $n$. Due to this specific resonance $\omega_{lib}=n$, this specific value $\sqrt{2}$ is suspected to be the limiting value of the secondary's shape beyond which the phase space of the secondary's rotational motion is dominated by chaos \citep{cuk2010orbital}. However, since $\mu$ can never be really zero in the BAS, this resonance $\omega_1=\omega_2$ can never happen. From the viewpoint of pure dynamics, this does not exclude the existence of the synchronous BAS consisting of a secondary with $a_b/b_B>\sqrt{2}$. We will come back to this at subsection \ref{sec:StableRegion}.

	\subsection{Periodic Orbits}
	\label{sec:PeriodicOrbits}
\begin{figure}
			\includegraphics[width=\columnwidth]{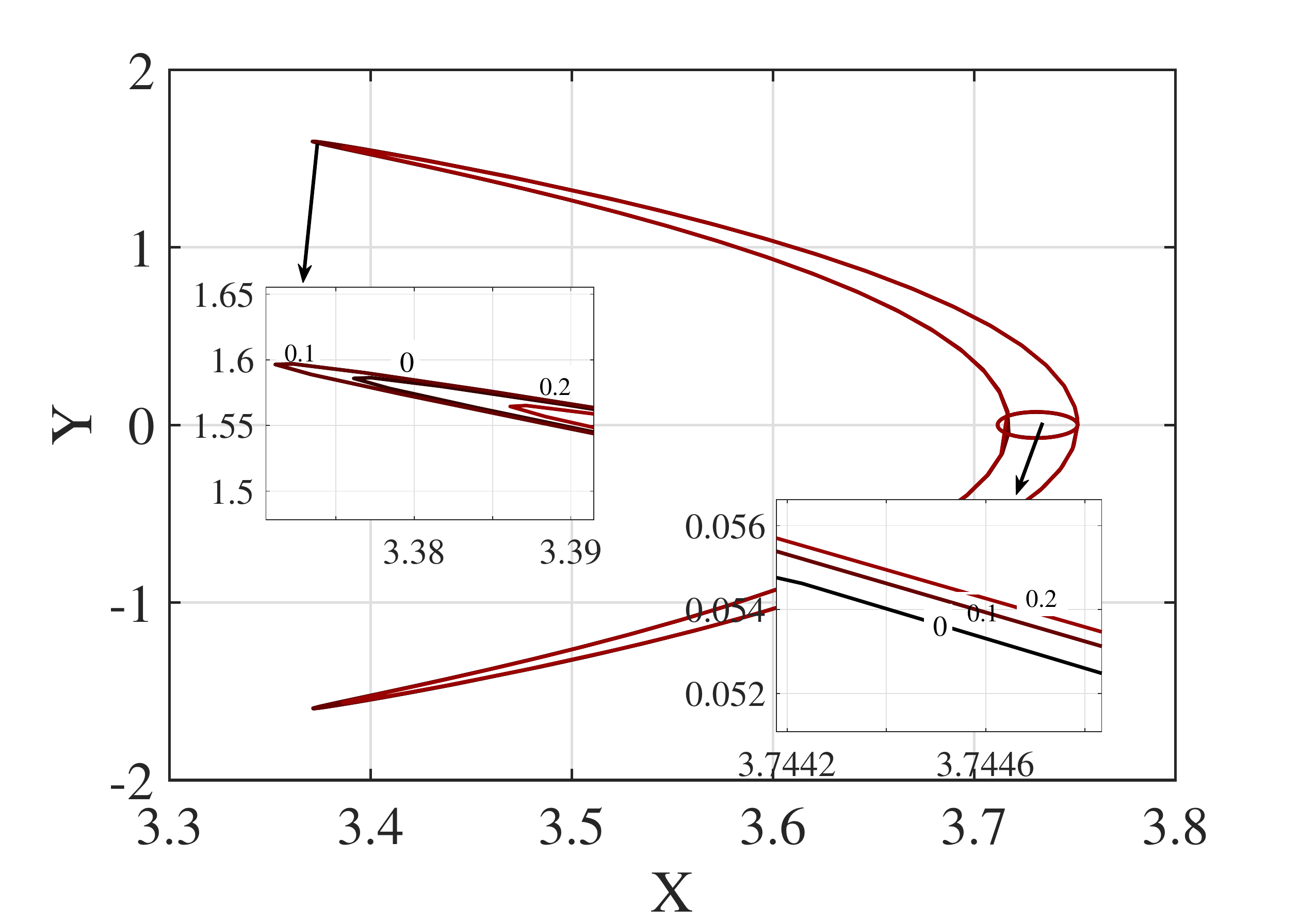}
			\caption{The long- and the short- period orbit with different $J_2^A$. The small boxes are local magnification of the orbits.}
			\label{fig:PeriodOrbit}
		\end{figure}

		Same as the sphere-ellipsoid model, the stable long-axis mode of the averaged ellipsoid-ellipsoid model has two kinds of period orbits around equilibrium. They are long-period and short-period orbits, and indicate two free components of the librational motion of the secondary's synchronous state. Introducing $\xi=r-r_0,\eta=\theta-\theta_0$, expanding Eq.\ref{eq:AveragedEOM_K} around EP and only retaining the linear terms, we have,
		\begin{equation}
			\left\{\begin{array}{l}
			\ddot{\xi}=a_{31} \xi+a_{34} \dot{\eta} \\
			\ddot{\eta}=a_{42} \eta+a_{43} \xi
			\end{array}\right.
		\label{eq:LinearedEquation}
		\end{equation}
		Solution to this linearized equation of motion takes the form of \citep{wang2020secondary}
		\begin{equation}
			\left\{\begin{array}{l}
			\xi=\alpha \cos \left(\omega_{2} t+\phi_{2}\right)+\frac{\beta}{a_{1}} \cos \left(\omega_{1} t+\phi_{1}\right) \\
			\eta=-a_{2} \alpha \sin \left(\omega_{2} t+\phi_{2}\right)-\beta \sin \left(\omega_{1} t+\phi_{1}\right) \\
			\xi=-\omega_{2} \alpha \sin \left(\omega_{2} t+\phi_{2}\right)-\frac{\beta \omega_{1}}{a_{1}} \sin \left(\omega_{1} t+\phi_{1}\right) \\
			\dot{\eta}=-\omega_{2} a_{2} \alpha \cos \left(\omega_{2} t+\phi_{2}\right)-\omega_{1} \beta \cos \left(\omega_{1} t+\phi_{1}\right)
			\end{array}\right.
		\label{eq:LinearedSolution}
		\end{equation}
		where, $\omega_1$ and $\omega_2$ are already given above. $\alpha$ and $\beta$ are two free parameters indicating the amplitudes of the long-period component and the short-period component. $\phi_1$ and $\phi_2$ are two free phase parameters. $\alpha_1$ and $\alpha_2$ are functions of the BAS parameters. Setting $\alpha=0$ in Eq.\ref{eq:LinearedSolution}, we get the linear approximation of the long-period orbit, and setting $\beta=0$, we get the linear approximation of the short-period orbit. Starting from these linear approximations we can generate the short-period and the long-period family using the well-known predict-correct algorithm. Details on how we generate these periodic orbits are omitted here. Readers can refer to \citep{wang2020secondary} and references therein. To show the difference between the averaged ellipsoid-ellipsoid model and the sphere-ellipsoid model, Fig.\ref{fig:PeriodOrbit} shows some examples of the short-period and the long-period orbits for different values of $J_2^A$. The small ellipses are the short-period orbits and the large elongated curves are the long-period orbits. Local magnifications of the curves are presented. Different curves correspond to different values of $J_2^A$. Again, the primary's shape has little influence on the periodic orbits' shape.

	\subsection{Stable Region}
	\label{sec:StableRegion}
\begin{figure}
			\begin{minipage}[t]{0.5\linewidth}
				\centering
				\includegraphics[width=\columnwidth]{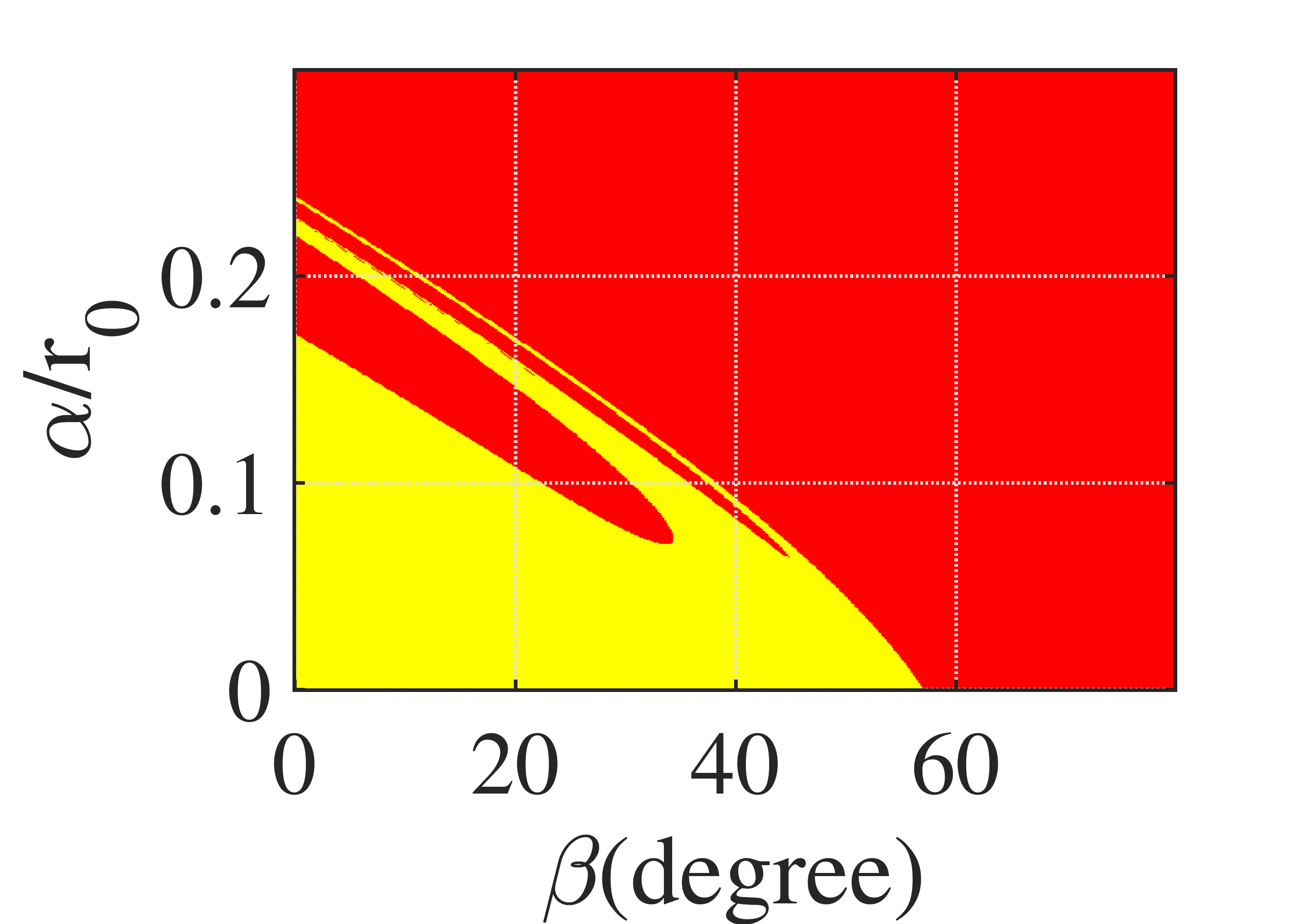}
			\end{minipage}%
			\begin{minipage}[t]{0.5\linewidth}
				\centering
				\includegraphics[width=\columnwidth]{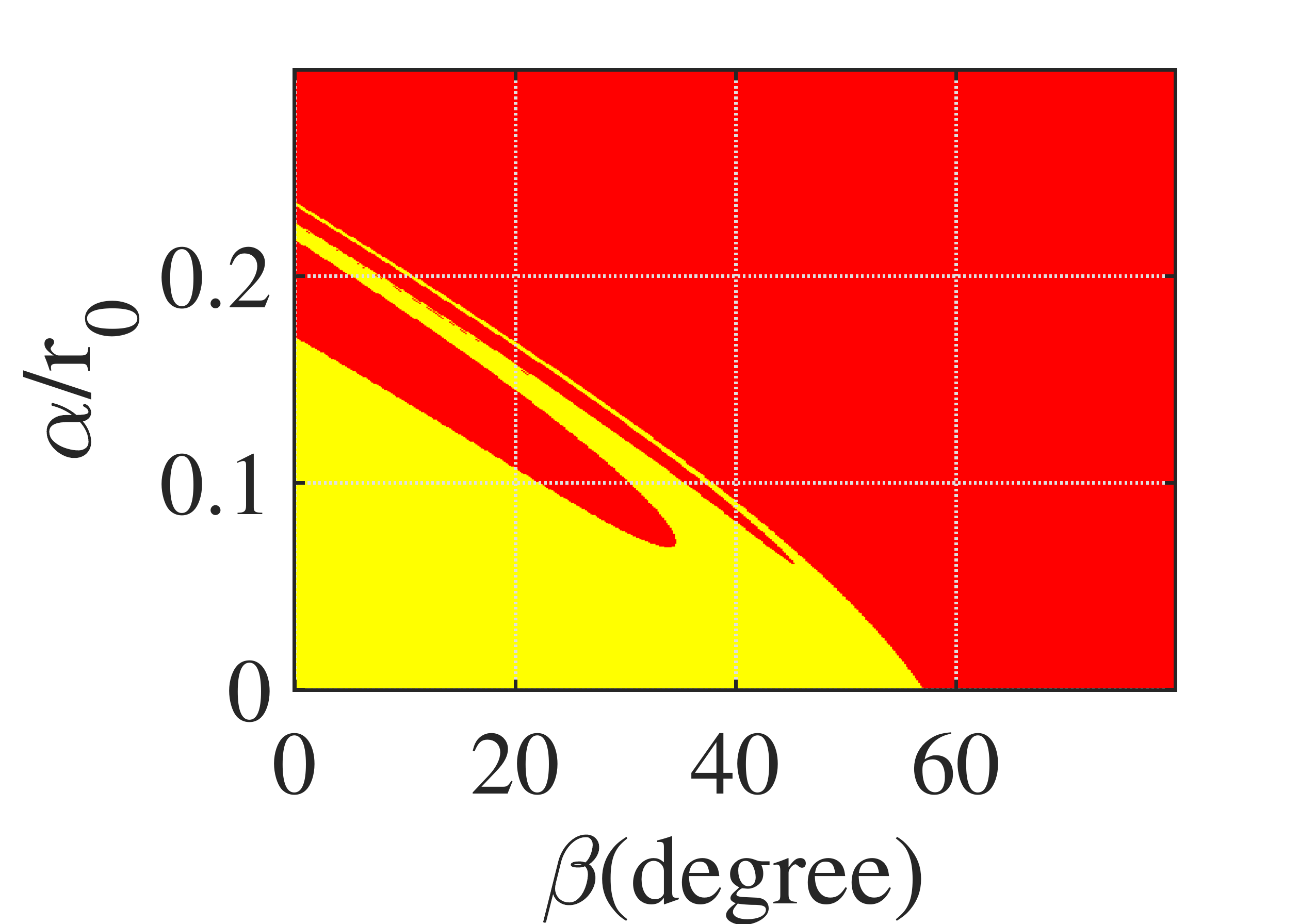}
			\end{minipage}
			\caption{Stable region of the secondary's synchronous state. Initial state vectors are provided by Eq.\ref{eq:LinearedSolution} with $\phi_1=\phi_2=0$. The absicissa $\beta$ is the amplitude of the long-period component, which indicates the libration amplitude of the 1:1 synchronous state. The ordinate $\alpha$ indicates the amplitude of the short-period component and $\alpha/r_0$ is an indicator of the orbit eccentricity. Yellow denotes the stable region and red is unstable region. The left is the sphere-ellipsoid model's stable region and the right is the averaged ellipsoid-ellipsoid model's stable region. Integration time of both two phase contours is $T_{int} = 150$ (dimensionless time units).}
		\label{fig:StablePic}
		\end{figure}

		Eq.\ref{eq:LinearedSolution} indicates that there are two free libration components of the motion around the equilibrium points. In the body-fixed frame of the secondary, the primary's trajectory appears as quasi-periodic trajectories with two free libraitonal frequencies $\omega_1$ and $\omega_2$. The secondary is trapped in the synchronous state as long as quasi-periodic orbits satisfy the condition that $|\eta|$ is smaller than $180^{\circ}$. Although it is only solution to the linearized model, Eq.\ref{eq:LinearedSolution} builds a 1-1 mapping between the initial state vector ($x_0,y_0,\dot{x}_0,\dot{y}_0$) at $t=0$ and the parameters ($\alpha,\beta,\phi_1, \phi_2$). Moreover, we set $\phi_1=\phi_2\equiv 0$. As a result, we have 
		\begin{equation} 
			\left\lbrace \begin{array}{ll} \xi_0=\alpha+\frac{\beta}{a_1}\\ \dot{\eta}_0=-\omega_2a_2\alpha-\omega_1\beta 
			\end{array}\right. 
		\end{equation} 
		This requires that the initial state vector lies at the $x$ axis, with an initial velocity perpendicular with the $x$ axis, i.e., $y=\dot{x}=0$. The above relation builds a 1-1 mapping between the initial state vector $x_0,\dot{y}_0$ and the two amplitude parameters $\alpha,\beta$. By choosing different values of $\alpha$ and $\beta$, we actually choose different trajectories. We integrate these trajectories for some time $T_{int}$(in our work, a value of 150 dimensionless units is used). If the synchronous state is broken within this time, it means that the combination of $\alpha$ and $\beta$ is unstable. If the synchronous state is preserved after this time, it means that the combination of $\alpha$ and $\beta$ corresponds to a stable synchronous state. By surveying the combinations of $\alpha$ and $\beta$, we are able to describe the stable region in the $\alpha-\beta$ plane. This technique was already used to generate the stable region for the sphere-ellipsoid model in our previous work. Here, we use the same technique to generate the stable region for the averaged ellipsoid-ellipsoid model. The results are shown in Fig.\ref{fig:StablePic}. The shape parameter for $A$ is $a_A/b_A = b_A/c_A = 1.2$. For comparison, the result of the sphere-ellipsoid model is also shown by fixing the mutual orbit distance and the mass fraction the same. It seems that the stable region of the secondary's synchronous state in the averaged ellipsoid-ellipsoid model has no obvious difference from that in the sphere-ellipsoid model. This means that the primary's non-spherical terms influence the secondary's synchronous state little.

		In the conventional model for the spin-orbit coupling, the mutual orbit is assumed invariant and the rotational motion of the satellite is one degree-of-freedom. When the satellite is trapped in the synchronous state, we denote the libration frequency as $\omega_{lib}$. Meanwhile, there is a forced component of the satellite's rotational motion induced by the mutual orbit eccentricity \citep{Murray1999solar}. In our work, the model described by Eq.\ref{eq:AveragedEOM_K} is a two degree-of-freedom system, and the there are two basic frequencies $\omega_1$ and $\omega_2$. In our previous work (see subsection 5.1 of that paper), we showed that in the limiting case ($\mu\rightarrow 0$) $\omega_1$ was actually the libration frequency $\omega_{lib}$ in the conventional model, and $\omega_2$ was actually the orbital frequency of the mutual orbit. Also, we showed that the parameter $\beta$ could be interpreted as the libration amplitude of the secondary's synchronous state and the parameter $\alpha/r_0$ could be interpreted as the mutual orbit eccentricity of the BAS. In Fig.\ref{fig:StablePic}, the result indicate that both sphere-ellipsoid and averaged ellipsoid-ellipsoid model bear the feature that there is an anti-correlation between the maximum orbit eccentricity and the libration amplitude of the synchronous state. More specific, according to Fig. \ref{fig:StablePic}, for a fixed value of $\beta$, there is a critical value of $\alpha/r_0$ beyond which the synchronous state of the BAS is broken. This critical value of $\alpha/r_0$ reduces when the value of $\beta$ increases, which means the orbit eccentricity of a BAS tends to be small if it has a large free libration amplitude. This phenomenon is already pointed out in our previous work and may help explain the observation fact that orbit eccentricity of the synchronous BAS tends to be small and the eccentric orbit usually be found in asynchronous BAS \citep{Pravec2016binary}.

		\begin{figure}
			\includegraphics[width=\columnwidth]{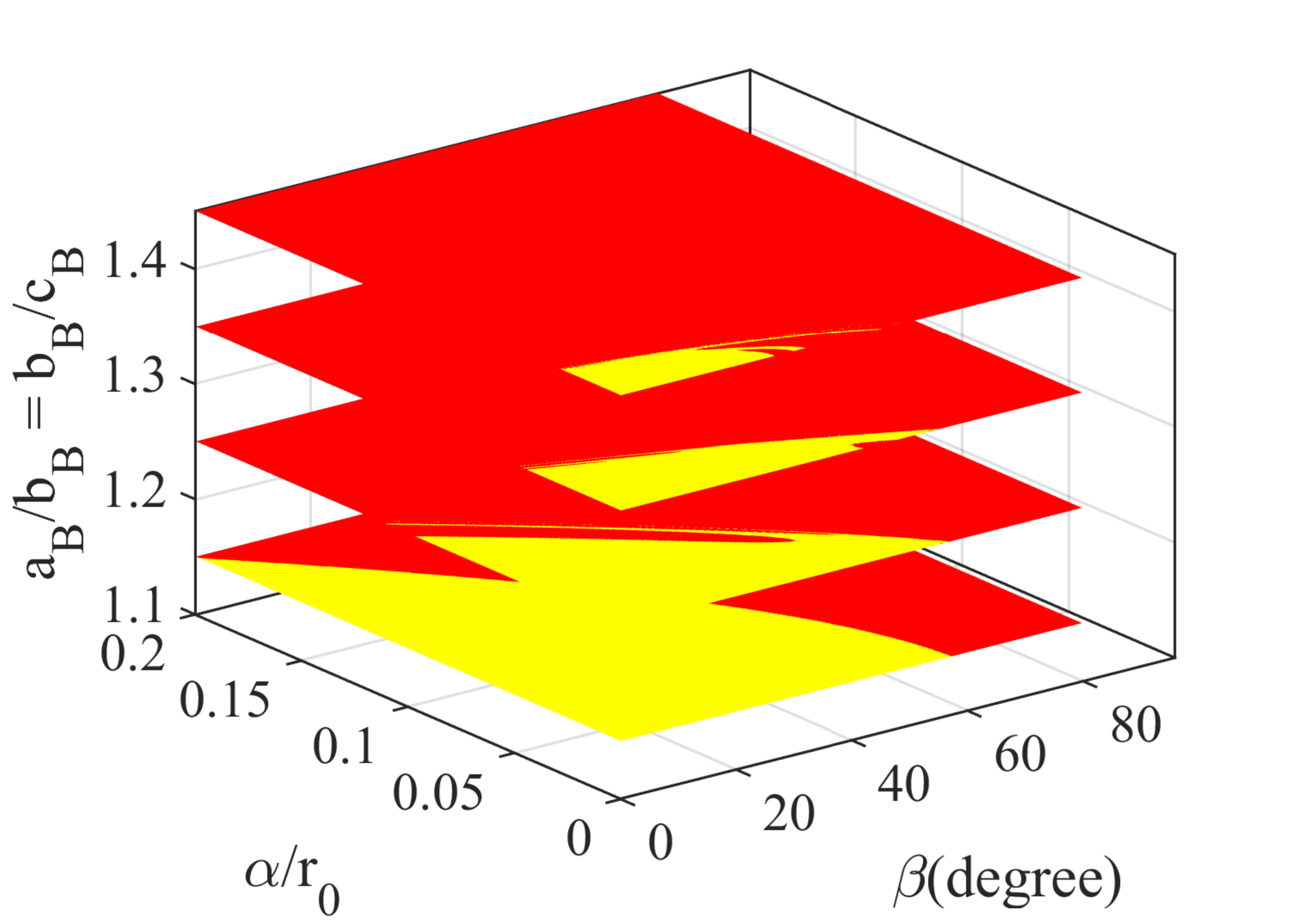}
			\caption{A stack of stable region contours is shown for different values of the secondary elongation $a_B/b_B$. From bottom to top, stable region shrinks with increasing $a_B/b_B$ in which the integration time is $T_{int} = 150$ (dimensionless time units).}
			\label{fig:SliceSecondaryElongation}
		\end{figure}

		In the current population of observed BASs, it seems that the shape parameter of the secondary $a_B/b_B$ is generally smaller than 1.5 \citep{Pravec2016binary}. We wonder whether there are some dynamics behind this phenomenon. \citep{cuk2010orbital} show in their numerical simulations that the secondary's rotational motion tends to be chaotic if $a_B/b_B>1.5$ and the mutual orbit eccentricity is larger than 0.05. Fixing the mass fraction, the mutual orbit distance, and the primary's shape, Fig.\ref{fig:SliceSecondaryElongation} shows the stable region for different values of $a_B/b_B$. An obvious fact is that the stable region shrinks with increasing value of $a_B/b_B$. At the value of $a_B/b_B=\sqrt{2}$ (which is close to 1.5), the maximum orbit eccentricity (here interpreted as $\alpha/r_0$) is about 0.05. This result agrees with the finding in \citep{cuk2010orbital}. Results in Fig.\ref{fig:SliceSecondaryElongation} do not completely rule out the existence of very elongated secondaries in the BASs. Such BASs may still exist as long as their mutual orbit eccentricity is small enough. The absence of very elongated secondaries in the current observed BAS population may be simply due to the lack of observation data. However, the weaker structural stability \citep{Sharma2014Stability} and stronger spin-orbit coupling \citep{Hou2017A} of elongated bodies may cause them easier to undergo secondary fission during the formation process of the BAS \citep{Jacobson2011dynamics}. This mechanism may limit the number of the BASs with highly elongated secondaries at the beginning of their formation \citep{Pravec2016binary}.

\section{Breakup Mechanism I --- Orbit Migration}
\label{sec:BreakupMechanismI}
	The above section is a continuation of our previous paper. The averaged ellipsoid-ellipsoid model shows that the primary's non-spherical terms have little influence on the secondary's rotational state, in terms of equilibrium points, periodic orbits, and stable region. In following sections, we focus on the main purpose of the current study---the breakup mechanism of the synchronous state of the BAS. Two mechanism will be studied in total. One is the orbit migration process caused by the BYORP torque and the tidal torques, and the other one is the spin-orbit resonance of the primary. In this section, we will study the migration process. For this purpose, the BYORP torque and the tidal torques should be introduced to the averaged equation, so Eq.\ref{eq:AveragedEOM} will be used.

	Eq.\ref{eq:AveragedEOM} is a 3-DOF dynamical system. Due to the presence of the BYORP torque, the total angular momentum is no longer a constant. Starting from a set of initial conditions, we can simulate the migration process of the synchronous BAS by propagating Eq.\ref{eq:AveragedEOM}. Some physical parameters of the BAS are displayed in Table \ref{tab:SystemParameters}. Depending on the direction of the BYORP torque, two kinds of migration process are available, inwards migration and outwards migration.

	The purpose of the study in this section is to show the migration process of the BAS in the stable contours as those displayed in Fig.\ref{fig:StablePic}. With the BAS migrating, we can expect that the BAS gradually approaches the boundary separating the stable and unstable regions. The problem is that Eq.\ref{eq:AveragedEOM} is no-longer angular momentum preserved due to the BYORP torque, so the 3-DOF system cannot be reduced to a 2-DOF by using the conservation of the total angular momentum. To solve this problem, the following technique is used. During the migration process of the BAS, at a specific epoch t, we have a set of system parameters
	\begin{equation}
		r(t),\Theta(t),\theta_B(t),\dot{r}(t),\dot{\Theta}(t),\dot{\theta}_B(t)
		\label{eq:ElementSix}
	\end{equation}
	The total angular momentum $K(t)$ of the BAS at this specific epoch can be calculated from above values. At this specific epoch $t$, we deliberately shut down the BYORP torque and the tidal torques. So the angular momentum is preserved and we can use the method described in section \ref{sec:DynamicsOfTheAveragedSystem} to generate stable region contour for this specific value of angular momentum. More details are described as follows.

	First, the state of the BAS is up to the values of Eq.\ref{eq:ElementSix}. We change the state vector of the BAS in Eq.\ref{eq:ElementSix} to the body-fixed frame of the secondary
	\begin{equation}
		r(t),\theta(t),\theta_B(t),\dot{r}(t),\dot{\theta}(t),\dot{\theta}_B(t)
		\label{eq:ElementSix_bodyfixed}
	\end{equation}
	And we change the equations of motion from Eq.\ref{eq:AveragedEOM} (by shutting down the BYORP torque and tidal torques) to Eq.\ref{eq:AveragedEOM_bodyfixed}. Then, by using the total angular momentum $K(t)$, we change from Eq.\ref{eq:AveragedEOM_bodyfixed} to the 2-DOF system Eq.\ref{eq:AveragedEOM_K}. We can calculate the equilibrium points of Eq.\ref{eq:AveragedEOM_K} for this specific value of $K(t)$, and the stable region by surveying different combinations of $\alpha$ and $\beta$ in Eq.\ref{eq:LinearedSolution} and setting $\phi_1=\phi_2=0$. At last, we transform the state vector of the BAS from Eq.\ref{eq:ElementSix_bodyfixed} to the state vector centered at the equilibrium point
	\begin{equation}
		\xi(t),\eta(t),\dot{\xi}(t),\dot{\eta}(t)
		\label{eq:Element_Linearization}
	\end{equation}
	By using the 1-1 mapping given by Eq.\ref{eq:LinearedSolution}, we can generate a set of parameters
	\begin{equation}
		\alpha(t),\beta(t),\phi_1(t),\phi_2(t)
		\label{eq:Element_Solution_Parameters}
	\end{equation}
	Usually $\phi_1(t)$ and $\phi_2(t)$ do not exactly equal zero. Since the stable region contour is generated by setting $\phi_1=\phi_2=0$, it is inappropriate to directly display $\alpha(t),\beta(t)$ in the stable region contour. To solve this problem, starting from the initial conditions given by Eq.\ref{eq:Element_Linearization}, we integrate Eq.\ref{eq:AveragedEOM_K} over a period of time and fix the value of $\dot{\xi}=0$. We can obtain a series of state vectors and transfer the state vector along the trajectory to $\alpha,\beta,\phi_1,\phi_2$ by using Eq.\ref{eq:LinearedSolution}. Then, we choose the $\alpha,\beta$ values with minimum $|\phi_1|+|\phi_2|$ values (usually they are very close to zero) and display them in the stable region contour. In this way, we are able to describe the position of the BAS in the stable region for this specific total angular momentum $K(t)$.

	Obviously, at different epochs of the migration process, the total angular momentum $K(t)$ is different and the stable region also changes. By choosing a series of epochs, we can view the BAS’ migration process in the stable region contour, like watching animations. By doing so, we actually take advantage of the fact that the period of the libration motion for the secondary is much shorter when compared with the migration process. In this following, we will use this technique to study the outwards and inwards migration process. In both cases, we focus on the difference between our model and the conventional model which decouples the orbital motion from the rotational motion. 

	\subsection{Outwards Migration}
	\label{sec:OutwardsMigration}
		During the outwards migration process, both BYORP torque and tidal torque increase the semimajor axis. Tidal torque is composed of two parts, $\ddot{\theta}_A^{tid}$ and $\ddot{\theta}_B^{tid}$. Usually the tidal torque in the BAS expands the orbit size. Also we choose a positive BYORP torque to expand the orbit size.

\begin{figure*}
			\includegraphics[width=2\columnwidth]{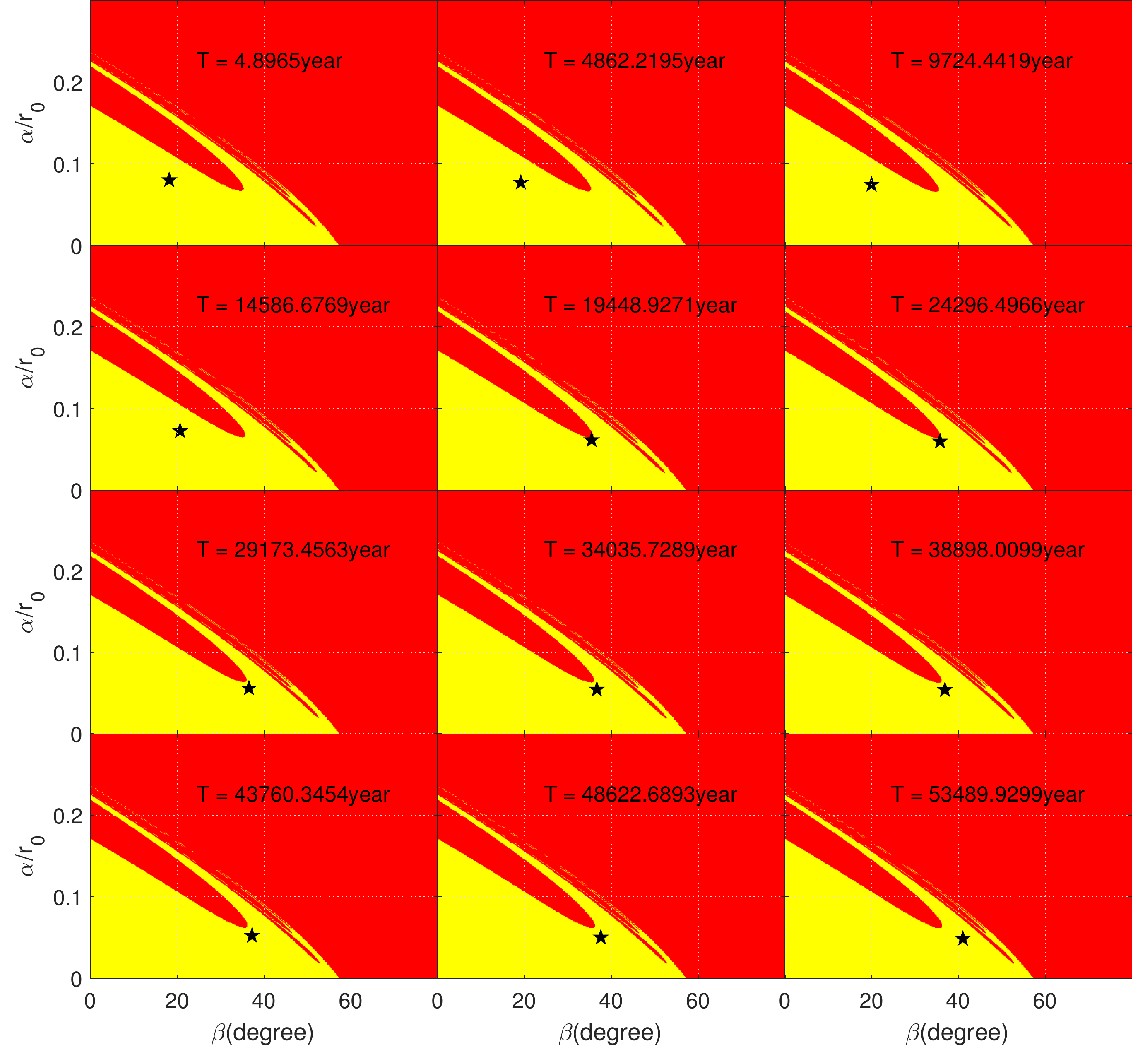}
			\caption{The process of outwards migration is shown in stable contour. The initial values are $\alpha/r_0 = 0.08,\beta=0$. The migration process agrees with the 1-DOF adiabatic invariance theory, in the sense that the orbit eccentricity (here the short-period component $\alpha/r_0$) decreases and the free libration amplitude (here the long-period component $\beta$) increases.}
			\label{fig:JointOut3}
		\end{figure*}

		Propagating the 3-DOF EOM including external torques and using the map technique introduced above, we can show the overall process of outwards migration in stable contour. The initial values $\alpha/r_0 = 0.08,\beta=0,\phi_1=\phi_2=0$ are used. Eq.\ref{eq:LinearedSolution} can transfer $\alpha,\beta,\phi_1,\phi_2$ to $r = r_0 + \xi,\theta = \eta,\dot{r} = \dot{\xi},\dot{\theta} = \dot{\eta}$. The libraitonal angel $\theta$ is criterion. If $\theta > 180 \degree$ during integration, we stop the integration and consider the synchronous state breakup. Throughout the whole integration, totally twelves shots of the migration process are taken and displayed in Fig.\ref{fig:JointOut3}.

		As we have mentioned previously, the ordinate $\alpha/r$ is an indicator of the orbit eccentricity and the abscissa is an indicator of the free libration amplitude. For the outwards migration, with the time increasing, the orbit eccentricity gradually reduces and the free libration amplitude gradually increases. The star displayed on the stable region gradually approaches towards right. Finally, the synchronous state breaks up when the free libration amplitude exceeds the stable region. This scenario agrees with the adiabatic invariance theory \citep{Jacobson2014formation}. The reason for this agreement is that the secondary is far from the primary in the outwards migration case. This leads to negligible effects of the secondary's rotational motion on the orbital motion. As a result, it is reasonable to decouple the orbital motion from the rotational motion. One remark is that the orbit eccentricity decreases in the outwards migration, so it cannot exceed its initial orbit eccentricity which is bounded by the stable region. As a result, wide asynchronous binary with a very large orbit eccentricity (such as 1998 ST27 listed in Table \ref{tab:SystemParameters} of \citep{Jacobson2014formation}) may not form by the outwards migration process.
		
	\subsection{Inwards Migration}
	\label{sec:InwardsMigration}
		According to Eq.\ref{eq:BYORPElementsEvolution}, when the BAS influenced by a negative BYORP torque, the orbit semi-major axis reduces and the orbit eccentricity increases. On the other hand, the two parts of tidal torque have opposite effects on the orbit eccentricity. Tides raised on the primary by the secondary excite the eccentricity, while tides raised on the secondary by the primary damp it. There are several possible outcomes: (1) The synchronous state is broken, and the BYORP process stops. The BAS migrates outwards again due to the tidal torques. According to the numerical simulations given by \citep{cuk2010orbital}, the secondary enters a chaotic rotation state when the synchronous state is broken and may be captured into the synchronous state again during the outwards migration, but with the other side facing the primary, causing the BYORP torque to flip its direction. In our analysis, we do find cases that the synchronous state can be broken by the inwards migration process, but we stop our investigation once the synchronous state breaks. Two examples will be given in subsection \ref{sec:TheBreakupOfSynchronousState}. (2) The tidal torque becomes stronger when migrating inwards. It is possible that the BYORP torque equals the tidal torques at a specific semi-major axis, leading to the long-equilibrium state of the BAS \citep{Jacobson2011Long}. In our study, we also find such cases. However, since we use the full-model simultaneously considering the orbital motion and the rotational motion, our study shows that the long-term state is never a real stable equilibrium state. The orbit eccentricity keeps increasing until finally goes out the stable region of the synchronous state. One example will be given in subsection \ref{sec:TheUnstableLong-termEquilibrium}. (3) The BAS keeps the synchronous state while migrating inwards until they collide again or are tidally disrupted \citep{cuk2010orbital}. In our simulations, a majority of the numerical simulations end into such a state where the long-term equilibrium behind the interior of primary. One example will be given in subsection \ref{sec:TheCollisionOfTwoBodies}.

		\subsubsection{The breakup of synchronous state}
		\label{sec:TheBreakupOfSynchronousState}
\begin{figure*}
				\includegraphics[width=2\columnwidth]{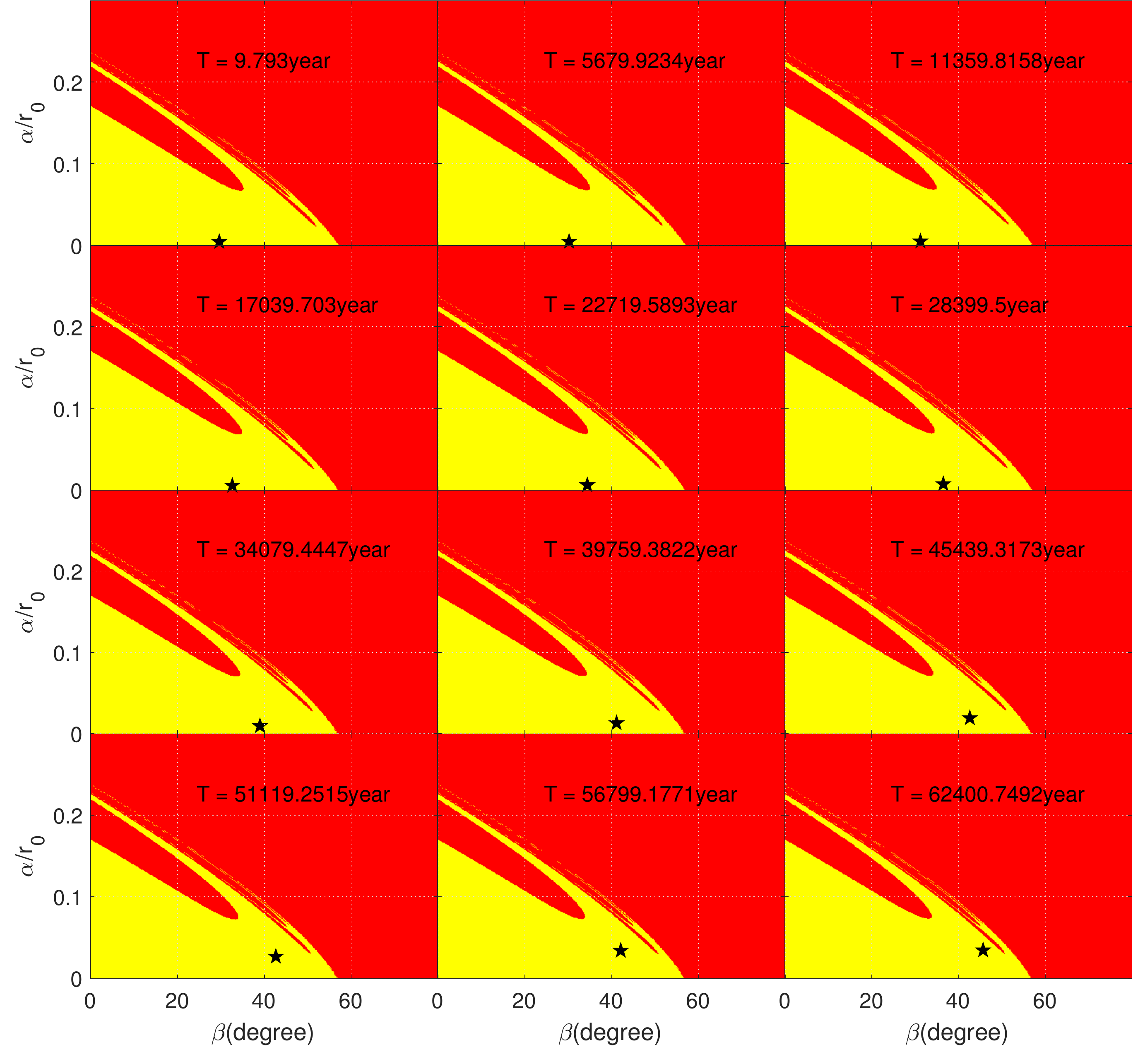}
				\caption{The process of inwards migration with initial values $\alpha/r_0=0$ and $\beta=30\degree$. During the migration process, the orbit eccentricity, i.e., the short-period component first remains small and then gradually increases. As for the 'free' libration amplitude, i.e., the long-period component, it gradually increases even in this inwards migration process. The secondary's synchronous state is broken up at a non-zero orbit eccentricity.}
				\label{fig:JointIn2}
			\end{figure*}

\begin{figure*}
				\includegraphics[width=2\columnwidth]{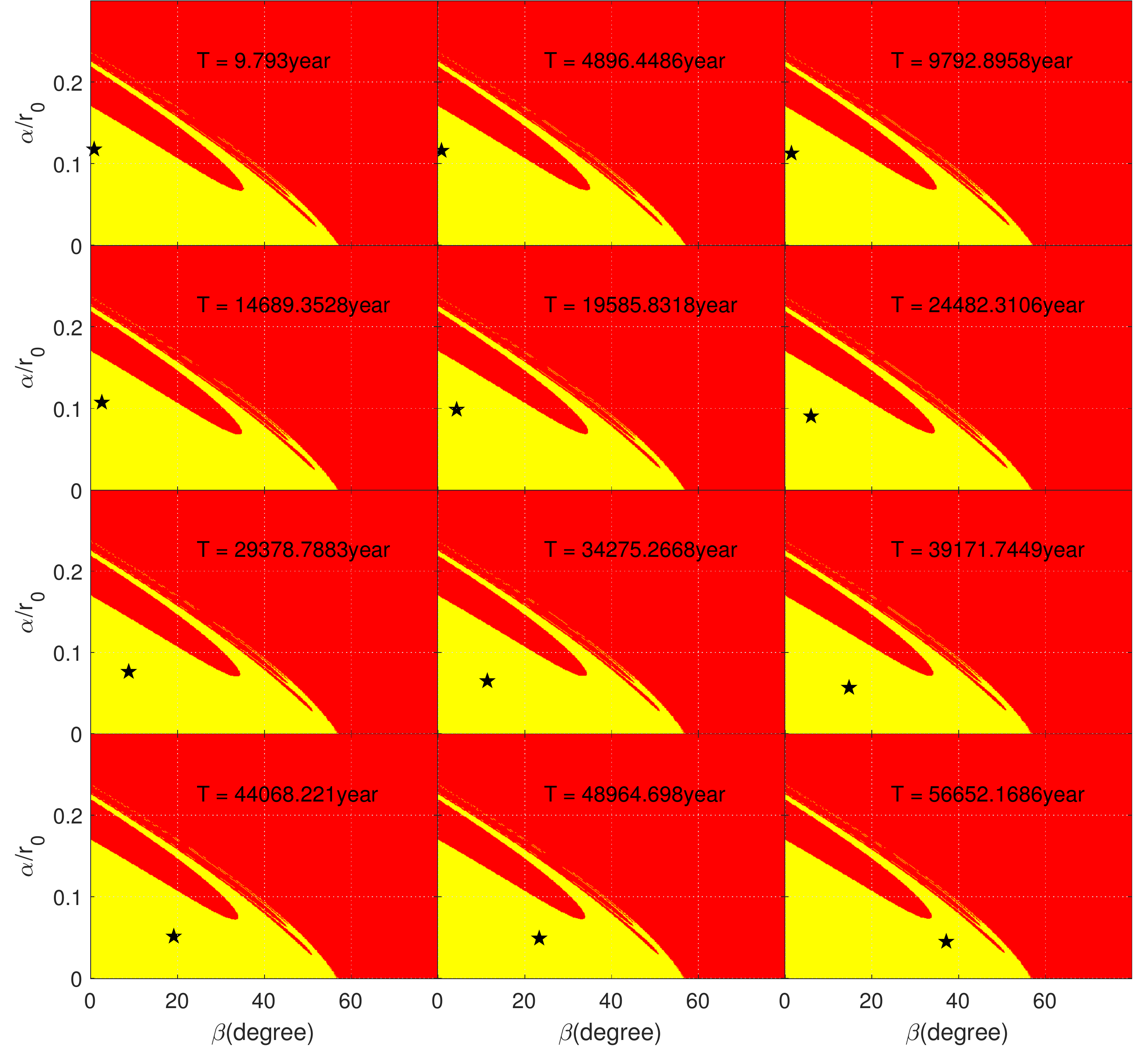}
				\caption{The process of inwards migration with initial values $\alpha/r_0=0.12$ and $\beta=0\degree$. In this numerical test, the orbit eccentricity (the short-period component) gradually decreases, while the 'free' libration amplitude (the long-period component) keeps growing until the secondary's synchronous state is broken.}
				\label{fig:JointIn1}
			\end{figure*}

			 In case of the inwards migration, according to the 1-DOF adiabatic invariance theory, the free libration amplitude should decrease due to the decreasing semi-major axis. However, our studies do not support this evolution senario. Two examples are shown in Fig.\ref{fig:JointIn2} and Fig.\ref{fig:JointIn1} respectively. For Fig.\ref{fig:JointIn2}, the initial configuration of the BAS is that the secondary has a large free libration component (i.e., a large long-period component $\beta=30\degree$) and a small orbit eccentricity (i.e., a small short-period component $\alpha/r_0=0$). When the orbit migrates inwards, the orbit eccentricity stays small at the first half of the migration but gradually increases at the latter half of the migration process. The free libration amplitude always increase during the whole inwards migration process. Finally, the synchronous state of the BAS is broken with a non-zero orbit eccentricity. For Fig.\ref{fig:JointIn1}, the initial configuration of the BAS is that the secondary has a small free libration component (i.e., a small long-period component $\beta=0$) but moves on an eccentric orbit (i.e., a large shot-period component $\alpha/r_0=0.12$). When the orbit migrates inwards, the orbit eccentricity gradually decreases, and the free libration amplitude gradually increases. Finally, the synchronous state breaks when the free libration amplitude exceeds the stable region. This is interesting, because in an inwards migration process, we find similar moving patten of the star in Fig.\ref{fig:JointIn1} as the one in Fig.\ref{fig:JointOut3} which is for the outwards migration process. An interesting phenomenon in both Fig.\ref{fig:JointIn2} and Fig.\ref{fig:JointIn1} is that the free libration amplitude also increases in the inwards migration process. This is similar to the outwards migration process. Our explanation for this phenomenon is that in the inwards migration process, the two asteroids are close to each other, and the spin-orbit coupling is strong. It is invalid to neglect the rotational motion's influence on the orbital motion and only considers the orbital motion's influence on the rotational motion \citep{Hou2017A}. Actually, in our work the free libration of the synchronous state is no longer 'free'. It is just a component of the synchronous motion in the full model. One remark is that the phenomenon shown in Fig.\ref{fig:JointIn2} and \ref{fig:JointIn1} may help explain the existence of asynchronous BAS with a close mutual distance and a small orbit eccentricity, for instance, (35107)1991 VH which is reported by \citep{Pravec2016binary}.

		\subsubsection{The unstable long-term equilibrium}
		\label{sec:TheUnstableLong-termEquilibrium}
			The tidal torques increases the orbit size and decrease the orbit eccentricity. The negative BYORP torque decreases the orbit size and increase the orbit eccentricity. According to \citep{Jacobson2011Long}, when the BYORP torque equals the tidal torque, the BAS may enter a long-term equilibrium state, as long as the changing rate of the orbit eccentricity is negative. Let $|F_B|=|F_T|$, the location of equilibrium is
			\begin{equation}
				r_e^7 = \frac{3k_A\mu_B^2\alpha_A^5}{2Q_A \pi B_s \alpha_B^2 f_{BY} H_s}.
			\label{eq:LongTermEquilibriumPoint}
			\end{equation}
			In which 
			$F_T = I_z^A \ddot{\theta}_A^{tid}/mr.$
			We set $\dot{\Theta} = \dot{\theta}_B$ so the tidal torque on $B$ is not considered. This equilibrium location depends on the tidal parameters and the BYORP coefficient directly. Using parameters in Tab.\ref{tab:SystemParameters}, we have $r_e = 2.2969$ agrees with our following simulation results. In \citep{Jacobson2011Long}, they used the first-order theory to find this long-term equilibrium too. During the migration process, if the mutual orbit distance is smaller the equilibrium value, dominate and the orbit expands. If mutual orbit distance is larger than the equilibrium value, the BYORP effect dominates and the orbit shrinks. In the 1-DOF adiabatic invariance theory, the effects of the secondary's rotational motion on the mutual orbital motion is actually neglected. As we have mentioned at the beginning of this work, we use the full model which simultaneously consider the orbital motion and the rotational motion. In this full model, we show that the long-equilibrium state of the BAS is actually unstable. 

\begin{figure*}
				\begin{minipage}[t]{0.5\linewidth}
					\centering
					\includegraphics[width=\columnwidth]{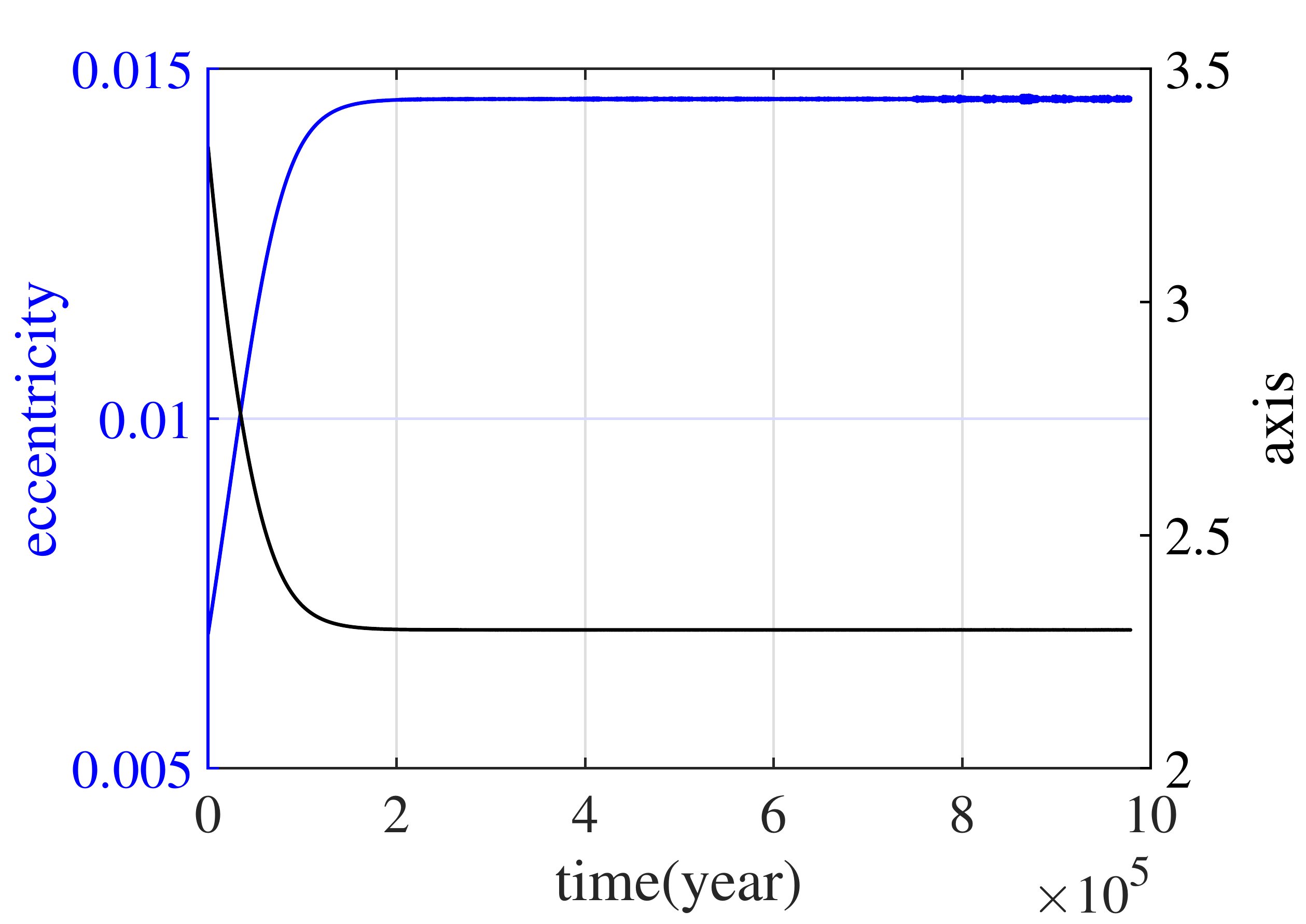}
				\end{minipage}%
				\begin{minipage}[t]{0.5\linewidth}
					\centering
					\includegraphics[width=\columnwidth]{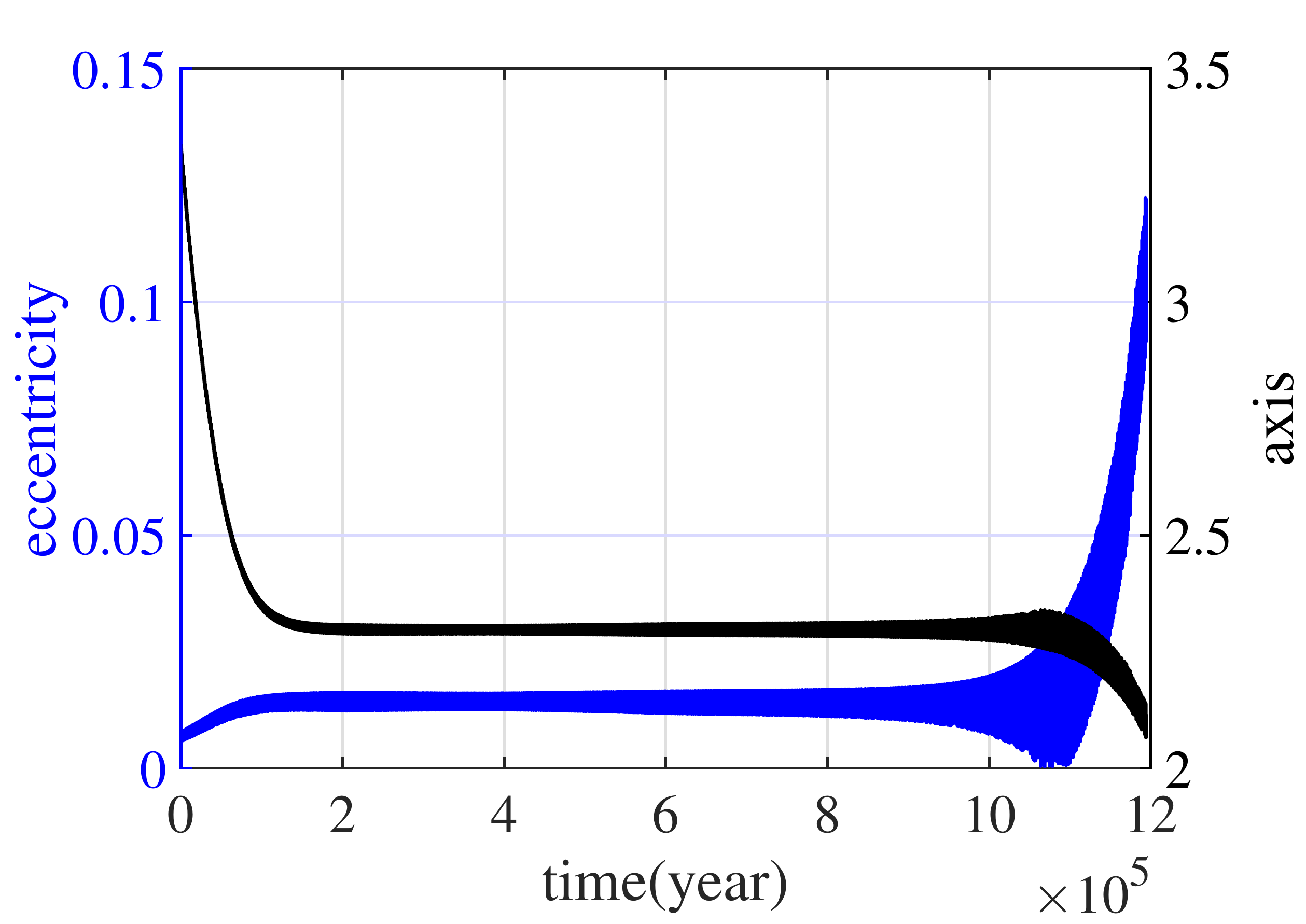}
				\end{minipage}
				\caption{Simulation results of the long-term stable equilibrium state between the BYORP torque and the tides. These two examples both start the inwards migration process at an initial mutual distance of $3.3$ but difference initial amplitudes ($\alpha=\beta=0$ for the left and $\alpha=0.001,\beta=10\degree$ for the right). For the left case, the BAS is trapped in the long-term equilibrium for more than $10^6$ years at a mutual distance of about $2.3$ which agrees with the computed value by the analytic formula Eq.\ref{eq:LongTermEquilibriumPoint}. In the right case which has a non-zero value of $\beta$, the BAS 'jumps' out of the long-term equilibrium state when the synchronous state is broken after about $10^6$ years.The timescale ($\sim 10^6$ years) for the orbit divergence agrees well with the analytical estimate $1/\gamma$.}
			\label{fig:Inwards8and15}
			\end{figure*}

            Supposing the BAS is already trapped in the long-term equilibrium state, the total angular momentum of the averaged ellipsoid-ellipsoid model plus tidal effects and BYORP effect described by Eq.\ref{eq:AveragedEOM} is conserved. Eq.\ref{eq:AveragedEOM} could be reduced by conservation of angular momentum $K$.  
			\begin{equation}
				\left\{\begin{array}{l}
				\ddot{r}=r\frac{(I_z^A\dot{\theta} - I_z^B\dot{\theta}-K)^2}{(mr^2 + I_z^B)^2}-\frac{1}{r^{2}}-\frac{3}{r^{4}}\left[A_1 + A_{2} \cos (2 \theta)\right] \\
				\ddot{\theta}=\frac{2\dot{r}(I_z^A \dot{\theta}_A - I_z^B \dot{\theta} - K)}{r(mr^2 + I_z^B)}-\frac{2A_{2} \sin (2 \theta) (I_z^B + mr^2)}{I_{z}^{B} r^{5}} - F_T - F_B
				\end{array}\right.
			\label{eq:AveragedEOM_K_FTB}
			\end{equation}
			Neglecting the terms $F_T$ and $F_B$, Eq.\ref{eq:AveragedEOM_K_FTB} is reduced to Eq.\ref{eq:AveragedEOM_K} which admits a stable equilibrium point corresponding to the exact synchronous state. Compared with other terms in Eq.\ref{eq:AveragedEOM_K_FTB}, the terms $F_T$ and $F_B$ are very small, and the equilibrium point of Eq.\ref{eq:AveragedEOM_K_FTB} should be very close to that of Eq.\ref{eq:AveragedEOM_K}. Expand Eq.\ref{eq:AveragedEOM_K_FTB} at this equilibrium point and we get the variation matrix of the linearized system, in the form of Eq.\ref{eq:AveragedEOM_K_FTB} at the long-term equilibrium point
			\begin{equation}
				A=\left(\begin{array}{cccc}
				0 & 0 & 1 & 0 \\
				0 & 0 & 0 & 1 \\
				a_{31} & a_{32} & 0 & a_{34} \\
				\overline{a}_{41} & a_{42} & a_{43} & a_{44}
				\end{array}\right).
				\label{eq:A44Perturbation}
			\end{equation}
			The matrix has a form which same as Eq.\ref{eq:A44Matrix}, but the element $\overline{a}_{41}$ has a different form
			\begin{equation}
				\overline{a}_{41} = a_{41}+\frac{B_{s} \pi \alpha_{B}^{2}f_{B Y}}{\left(a_{s} /[L]\right)^{2} \sqrt{1-e_{s}^{2}}}  \frac{\overline{F_{s}}}{m r^2} + \frac{7}{m r^2}\left[I_{z}^{A} \ddot{\theta}_{A}^{t i d}+I_{z}^{B} \ddot{\theta}_{B}^{t i d}\right].
				\label{eq:a41Overline}
			\end{equation}
			Due to this difference, eigenvalues of Eq.\ref{eq:A44Perturbation} has a following form
			$$\gamma \pm i \omega_{l} ; -\gamma \pm i \omega_{s}$$
			where $\gamma_*$ is a real number (for the case studied by us, $\gamma=9.224 \times 10^{-12}$). Solution to the linearized system no longer takes the form of Eq.\ref{eq:LinearedSolution}. It has the following form
			\begin{equation}
				\left\{\begin{array}{l}
				\xi=\alpha e^{-\gamma t} \cos \left(\omega_{s} t+\phi_{2}\right)+ \frac{\beta}{\alpha_{1}} e^{\gamma t} \cos \left(\omega_{l} t+\phi_{1}\right) \\
				\eta=- \alpha_{2} \alpha e^{-\gamma t} \sin \left(\omega_{s} t+\phi_{2}\right)- \beta e^{\gamma t} \sin \left(\omega_{l} t+\phi_{1}\right)
				\end{array}\right..
			\label{eq:LinearedSolution_Perturbation}
			\end{equation}
			Eq.\ref{eq:LinearedSolution_Perturbation} tells us that when the BAS is trapped in the long-term equilibrium state, the orbit eccentricity ($e \sim e^{-\gamma t} \alpha / r_0$) and the free libration amplitude ($\theta \sim e^{\gamma t} \beta$) both diverge exponentially. Although the value of $\gamma$ is small, judging from Eq.\ref{eq:LinearedSolution_Perturbation} that the amplitude of the long-period component increases exponentially due to the presence of the $e^{\gamma t}$ term. This means the BAS will eventually break up no matter how small the initial value of $\beta$ is as long as it does not exactly equal zero. Two examples are shown in Fig.\ref{fig:Inwards8and15}. The BAS (left) starts the inwards migration process at an initial mutual distance  $r_0=3.3$ and amplitudes $\alpha=0,\beta=0$. the BAS is trapped in the long-term equilibrium state for more than $10^6$ years at a mutual distance of about $2.3$ which agrees with the computed value by the analytic formula Eq.\ref{eq:LongTermEquilibriumPoint}. The BAS (right) starts the inwards migration process at a same initial mutual distance but different initial amplitudes $\alpha = 0.001,\beta=10 \degree$. The BAS is trapped in the long-term equilibrium state for a long time. After about $10^6$ years, due to the non-zero value of $\beta$ the secondary's synchronous eventually break up. Along with the breakup of the synchronous state is the disappearance of the BYORP torque and the long-term equilibrium state. The timescale ($\sim 10^6$ years) for the orbit divergence shown in Fig.\ref{fig:Inwards8and15} agrees well with the analytical estimate $1/\gamma$. One remark is, for the migration process shown in the left frame of Fig.\ref{fig:Inwards8and15}, the BAS will eventually break up after a longer time. This is because Eq.\ref{eq:LinearedSolution_Perturbation} is only a solution of the linearized system, so $\beta=0$ does not exactly mean a zero value of $\beta$ in the full model.

			Studies in this subsection indicates that the true long-term equilibrium state of the BAS may never exist, although the BAS can be temporarily captured in this state for quite a long time. The difference between our model and the model used by \citep{Jacobson2011Long} is that we simultaneously consider the orbital and the rotational motion during the migration process. The rotational motion of the secondary, has a remarkable influence (through the spin-orbit coupling mechanism) on the orbital motion after an exponential time when combined with the tidal effects and the BYORP effect at the long-term equilibrium state. The same result can also be interpreted as the 1:1 spin-orbit resonance (i.e., the synchronous state) of the secondary being gradually destroyed by dissipative tidal and BYORP effects, similar to the phenomenon that orbital resonances gradually destroyed by the thermal Yarkovsky effect \citep{Hou2016dynamics,wang2017dynamical}
			
		\subsubsection{The collision of two bodies}
		\label{sec:TheCollisionOfTwoBodies}
			In case that the equilibrium distance given by Eq.\ref{eq:LongTermEquilibriumPoint} is smaller than the sum of the two asteroids' sizes, the two asteroids may collide with each other as long as the synchronous state is not broken before collision. In our simulations, we do find such cases. No examples are given here.

\section{Breakup Mechanism II --- spin-orbit resonance of the Primary}
\label{sec:BreakupMechanismII}
	The primary's rotational motion is actually decoupled (by averaging) from the full model used in the above section which is dedicated to the synchronous state breakup problem caused by the migration process. In this section, we study the primary's rotational motion on the secondary's synchronous state. Generally, when $A$ is rotating fast, its direct and indirect effects on secondary's rotation \citep{Hou2017A,Hou2018note} is of short period and the above averaged ellipsoid-ellipsoid model is a good approximation. However, this scenario may be different when the primary's rotation crosses some major spin-orbit resonances. The secondary's synchronous state may be broken when the primary crosses these spin-orbit resonances.

\begin{figure*}
	    \begin{minipage}[t]{0.5\linewidth}
			\centering
			\includegraphics[width=\columnwidth]{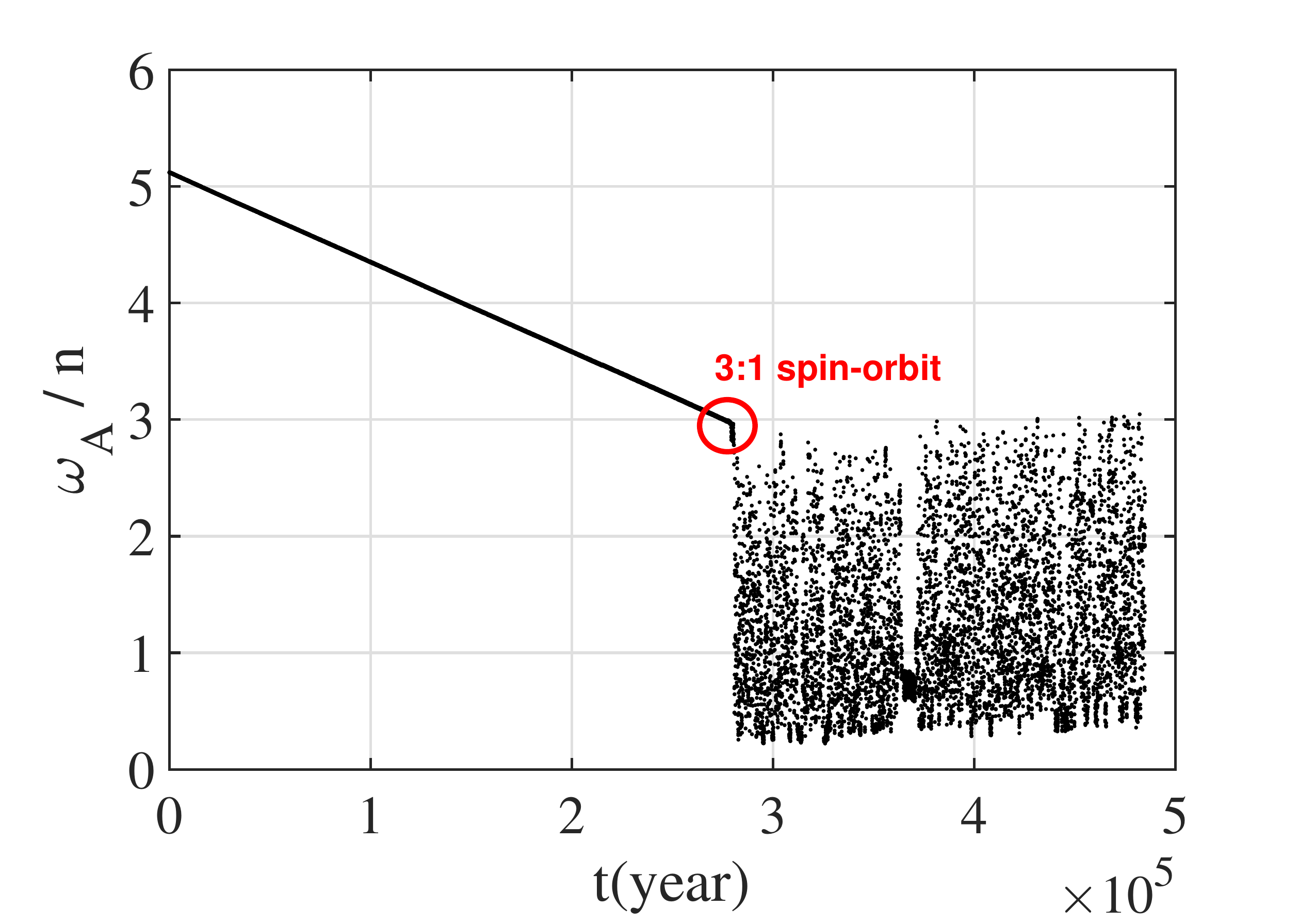}
		\end{minipage}%
		\begin{minipage}[t]{0.5\linewidth}
			\centering
			\includegraphics[width=\columnwidth]{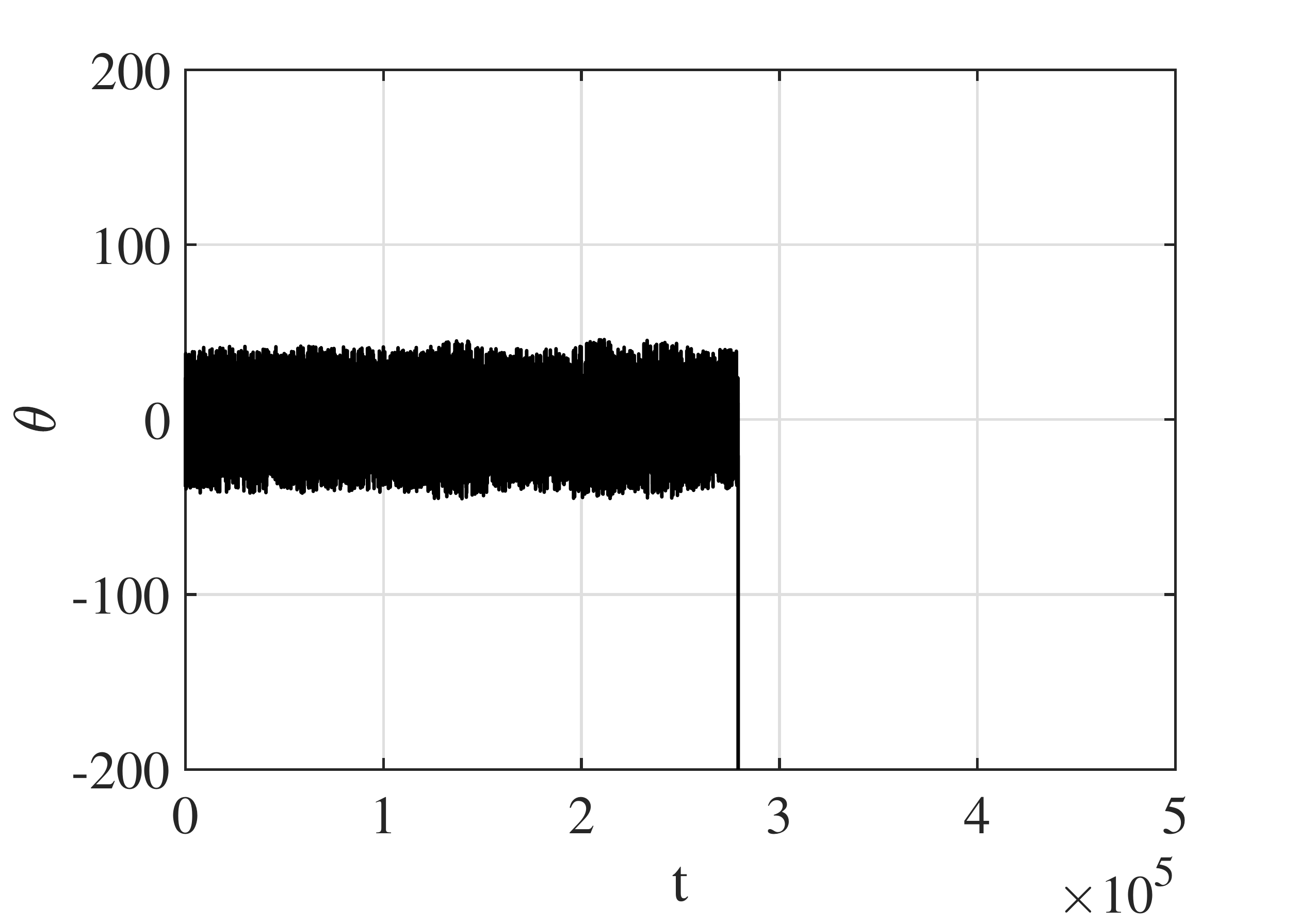}
		\end{minipage}
		\caption{Time history of the primary's rotation speed (left) and the libration angle of the secondary (right). The primary's rotation speed firstly uniformly reduces due to its negative YORP torque. When its rotation crosses the 3:1 spin-orbit resonance, it becomes chaotic. Through the spin-orbit-spin coupling mechanism, the secondary's rotation also becomes chaotic and its synchronous state is broken.}
	\label{fig:ResonanceA}
	\end{figure*}

	The primary's rotation state may be gradually reduced by the YORP effect or by the tidal effects. In this section, in order to remove the tidal effects and the BYORP effects, they are shut down from the force model. The equations of motion by Eq.\ref{eq:YHC4} for the ellipsoid-ellipsoid model is used. However, to simulate the gradual decrease of the primary's rotation speed, a negative YORP torque is added to $A$'s rotational motion. For the BAS given in Table \ref{tab:SystemParameters}, the torque is $$\ddot{\theta}_A^Y = \frac{\alpha_A}{1-\mu}\frac{B_s\overline{F_{s}}}{(a_s/[L])^2\sqrt{1-e_s^2}},$$ which is given by previous researches \citep{Rozitis2013strength,Rossi2009computing}. Time history curves of $A$'s rotational motion and the secondary's synchronous angle (the angle $\theta$ in Fig.\ref{fig:ModelAB}) are shown in Fig.\ref{fig:ResonanceA}.

	In our simulation displayed in Fig.\ref{fig:ResonanceA}, the migration starts from $\omega_A/n = 5.123$ where $n$ is the mutual orbit frequency. The spin-orbit resonance could be named as $\omega_A : n$ resonance. At the $\omega_A/n = 3$, the 3:1 spin-orbit resonance breaks up the secondary's synchronous state, and the orbital and rotational motions of the BAS become chaotic. We believe this phenomenon is closely associated with the strong spin-orbit and spin-orbit-spin coupling mechanism of the BAS \citep{Batygin2015Spin,Nadoushan2016Geography,Hou2017A}. $A$'s rotational motion influences $B$'s rotational motion through these mechanisms. When $A$ crosses the chaotic separatrix surrounding the major spin-orbit resonances \citep{Hou2017A,Naidu2015near}, its rotational motion becomes chaotic. Through the coupling, $B$'s rotational motion may also become chaotic and thus its synchronous state is broken. 
	
	A natural question is whether this type of breakup mechanism works at even higher order spin-orbit resonances. In our numerical simulations, however, we failed to find any breakup phenomenon at spin-orbit resonances with orders higher than 3:1. One possible explanation is that the strength of spin-orbit resonances with orders higher than 3:1 is too weak so that they can be easily crossed.
	
	In the current population of synchronous BAS, the primary generally rotates faster than 3 times the orbital frequency, but this does not exclude the possibility that a slowly rotating solitary asteroid may have a satellite in the past and lose it when primary crosses the major spin-orbit resonances. Because according to our numerical simulations this phenomenon does exist, although currently we are not sure to what extent this breakup mechanism works. Our next step is to find out evidence of this breakup mechanism in the current population of BAS or binary pairs. This will be the focus of our future study. One remark is: for this mechanism to work, the primary cannot rotate fast.
	
\section{Conclusion}
\label{sec:DiscussionandConclusion}
	This paper continues the authors' previous work \citep{wang2020secondary} and presents a coplanar averaged ellipsoid-ellipsoid model of synchronous binary asteroid system (BAS) plus thermal and tidal effects. This model is different from the classical spin-orbit coupling model which neglects the rotational motion’s influence on the orbital motion. The orbital and the rotational motions are considered simultaneously. In our study, the following findings are achieved.

	(1) In case that the primary rotates fast, its shape (non-spherical gravity) has negligible effects on the secondary's synchronous state. The stable region of the secondary's synchronous state is mainly determined by the secondary's shape. 
	
	(2) With the increase of the secondary's shape parameter $a_B/b_B$, stable region of the secondary's synchronous state shrinks continuously. When $a_B/b_B$ is larger than $\sqrt{2}$, the stable region of the secondary's synchronous state is small but not zero. The mutual orbit eccentricity of a synchronous BAS in this case generally should be smaller than $0.05$.

	(3) In the inwards migration, the breakup of the synchronous state could also be caused by the fact that the 'free' libration amplitude exceeds the limit.

	(4) Moreover, the long-term stable equilibrium state between the BYORP torque and the tides in the inwards migration process may eventually be broken up in our full model.  
		
\section*{Acknowledgements}

This work is supported by the National Natural Science Foundation of China(11773017, 11703013, 11673072).

\section*{Data Availability}
The data underlying this article are available in the article.

%%%%%%%%%%%%%%%%%%%%%%%%%%%%%%%%%%%%%%%%%%%%%%%%%%



\bibliographystyle{mnras}
\bibliography{example} % if your bibtex file is called example.bib

%\begin{thebibliography}{99}
%\bibitem[\protect\citepauthoryear{Author}{2012}]{Author2012}
%\bibitem[\protect\citepauthoryear{Others}{2013}]{Others2013}
%\end{thebibliography}

%%%%%%%%%%%%%%%%%%%%%%%%%%%%%%%%%%%%%%%%%%%%%%%%%%

%%%%%%%%%%%%%%%%% APPENDICES %%%%%%%%%%%%%%%%%%%%%

\appendix

\section{Elements of the matrix $A_{4 \times 4}$}
\label{sec:AppendA}
	\begin{equation}
		\begin{array}l
			a_{31} =\frac{(-I_z^B\dot{\theta}-K)^2(I_z^B - 3mr^2)}{(I_z^B+mr^2)^3}+\frac{2}{r_0^3}+6\frac{A_1+A_2\cos(2\theta_0)}{r_0^5}\\
			a_{32} =\frac{6A_2\sin(2\theta_0)}{r_0^4}\\
			a_{33} = 0\\
			a_{34} =\frac{2rI_z^B(K+I_z^B\dot{\theta})}{(I_z^B+mr^2)^2}\\
			a_{41} =\frac{2\dot{r}(I_z^B+3mr^2)(K+I_z^B\dot{\theta})}{(I_z^B+mr^2)^2r^2}+\frac{A_2(6mr_0^3+5I_z^B)\sin(2\theta_0)}{I_z^Br_0^6}\\
			a_{42} =-\frac{2(I_z^B + 2mr^2)A_2\cos(2\theta_0)}{I_z^Ar_0^5}\\
			a_{43} =-\frac{2(I_z^B\dot{\theta}_0+K)}{r_0(mr_0^2+I_z^B)}\\
			a_{44} =-\frac{2\dot{r}_0I_z^B}{(I_z^B+mr^2)r_0}
		\end{array}	\notag
	\label{eq:A44}
	\end{equation}
\bsp	% typesetting comment
\label{lastpage}
\end{document}